\documentclass[12pt, draftclsnofoot, onecolumn]{IEEEtran}

\usepackage{tipa}
\usepackage{amsmath}
\usepackage{mathrsfs}
\usepackage{dsfont}
\usepackage{cleveref}

\usepackage{graphicx}
\usepackage{subfigure}

\usepackage{multicol}
\usepackage{amssymb}
\usepackage{multirow}
\usepackage{longtable}
\usepackage[numbers,sort&compress]{natbib}
\usepackage{enumerate}

\usepackage{stfloats}
\usepackage{algorithm}
\usepackage{algorithmic}
\usepackage{multirow}
\usepackage{amsmath}
\usepackage{xcolor}

\usepackage{graphicx,times,amsmath,booktabs,algorithm,algorithmic,amsmath}
\usepackage{multicol}
\usepackage{amsmath,amssymb}
\usepackage{subfigure}
\usepackage{stfloats}
\usepackage{caption2}

\begin{document}

% paper title
\title{Collaborative Multi-Resource Allocation in Terrestrial-Satellite Network Towards 6G}
\author{
\IEEEauthorblockN{Shu Fu, Jie Gao, and Lian Zhao}
%\thanks{This work was supported in part by the National Natural Science Foundation of China under Grant 61701054; and in part by the Fundamental Research Funds for the Central University under Grant 2020CDJQY-A001.}
\thanks{Co-corresponding authors: Shu Fu and Lian Zhao.}
\thanks{Shu Fu is with the School of Microelectronics and Communication Engineering, Chongqing University, Chongqing, P. R. China, 400044. He is also with the State Key Laboratory of Integrated Services Networks, Xidian University, Xi'an, Shaanxi, 710071, P. R.  China. (e-mail: shufu@cqu.edu.cn).}
\thanks{Jie Gao is with the Department of Electrical and   Computer Engineering, Marquette University, Milwaukee, WI, 532331, USA (email: j.gao@marquette.edu).}
\thanks{Lian Zhao is with the Department of Electrical, Computer, and Biomedical Engineering, Ryerson University, Ontario, Canada, M5B 2K3 (e-mail: l5zhao@ryerson.ca).}
}
\maketitle
\begin{abstract}
Terrestrial-satellite networks are envisioned to play a significant role in the sixth-generation (6G) wireless networks.
%In order to improve the system energy efficiency in terrestrial-satellite communication, the resources of caching, computing, and communication (3C) should be  allocated in a collaborative manner in terrestrial-satellite networks (TSN).
In such networks, hot air balloons are useful as they can relay the signals between satellites and ground stations.
Most existing works assume that the hot air balloons are deployed at the same height with the same minimum elevation angle  to the satellites,
{\color{blue}which may not be practical due to possible route conflict with airplanes and other flight equipment.}
%Moreover, the considered system model in the existing works focus on only one part in TSN, lacking the viewpoint from the integrated system of ground stations, hot air balloons, and satellites.
In this paper, we consider a TSN containing hot air balloons at different heights and with different minimum elevation angles, which creates the challenge of non-uniform available serving time for the communication between the hot air balloons and the satellites.
{\color{blue}Jointly considering the caching, computing, and communication (3C) resource management for both the ground-balloon-satellite links and inter-satellite laser links, our  objective is to maximize the network energy efficiency.
Firstly,  {\color{blue} by proposing a tapped water-filling algorithm}, we schedule the traffic to relay among satellites according to the available serving time of satellites.
Then, we generate a series of configuration matrices, based on which we formulate the relationship of relay time and the power consumption involved in the relay among satellites.}
Finally, the integrated system model of TSN is built and solved by geometric programming with Taylor series approximation.
Simulation results demonstrate the effectiveness of our proposed scheme.
\end{abstract}

\section{Introduction}
\IEEEPARstart{I}{n} 6G, terrestrial-satellite networks (TSN) will become the key leverage to explore airspace resources for communications \cite{ADDINGSHU1, TVT2019CQ, PauloDeepTSN, YCLiang}.
In such networks, laser based inter-satellite links working at Terahertz frequency can support high-capacity transmissions among low earth orbit (LEO) satellites \cite{ZhisurveyTSN}.
Meanwhile, the large distance between satellites and ground stations necessitates relays, such as hot air balloons, between the terrestrial and satellite segments \cite{Zhang6GTSN}.
 A TSN integrates caching, computing, and communication (3C), each of which
may consume considerable energy.
Therefore, maximizing the energy efficiency performance will be one of the main design objectives of TSN in 6G \cite{TMC2018ShanZhang, Communication2018SF}.

Communications in a TSN involve steps of data traffic  collection, transmission, and relay.
%{\color{blue}The system delay in TSN can achieve a tradeoff between the resource utilization of the steps to maximize the system energy efficiency.}
In a previous work \cite{ADDINGSHU1}, we studied such a TSN with the above steps but without considering hot air balloons as relays or laser link among satellites.
{\color{blue}In practice, a TSN can be very complex due to the following reasons.}

{\color{blue} Firstly, there can be relays such as hot air balloons between satellites and ground stations. In real-world environment, the relays  may be deployed at different heights due to a variety of reasons, such as avoiding airplane routes.
Additionally, the minimum elevation angle from the hot air balloons to the satellites may not be identical either.
Define the time allowing communications between a relay and its serving satellite as \emph{time windows}.
Then, the time windows for each relay to communicate to a satellite can be unique.
Modeling and studying the above real-world TSN scenario is challenging and important.
However, relays with different heights and minimum elevation angles are not yet considered in the existing works, to the best of our knowledge.}

{\color{blue}Secondly, regarding the laser links among the satellites, it is necessary to determine the number of lasers per satellite as well as develop a proper scheme for scheduling inter-satellite traffic. If the number of lasers per satellite is  large, delay is low, power consumption can be high, and scheduling can be complex.
By contrast, if the number of lasers per satellite is small, power consumption is low, but the delay is large.}

{\color{blue}Finally, in additional to the energy efficiency of communications between the relays and  satellites and among the satellites, as mentioned above, the communications between  ground stations and hot air balloons should be involved while considering the network energy efficiency.
Moreover, the energy consumption of caching and computing should also be included, in additional to that of communication. This requires a joint allocation of the 3C resources.}

Existing works contributed to TSNs from different aspects.
In \cite{BohaoEnablingTSN}, a framework to efficiently deploy customized service function chains was proposed in terrestrial-satellite hybrid cloud networks, which enables computation and data traffic  off-loading in a TSN.
In \cite{CuiqinConstellationTSN}, a genetic algorithm based method was proposed to increase  the  coverage of a TSN.
In \cite{BoyaUltraTSN}, a novel network architecture was proposed to provide seamless and high data-rate wireless service in a TSN, where  the terrestrial and satellite segments are jointly designed.
In \cite{NingSoftwareTSN}, a dynamic space-air-ground resource pool was proposed by a software-defined space-air-ground integrated framework to effectively manage the network in a seamless, efficient, and cost-effective manner.
In \cite{JingjingUAVTSN}, an unmanned aerial vehicle (UAV) aided space-air-ground network was proposed to provide seamless coverage and high  system throughput.
However, the system models used in most of the existing works considered the case that all hot air balloons were deployed at an equal height and with the same minimum elevation angle.
In \cite{ShuUAVTSN}, a UAV based relay system was proposed in a non-terrestrial network.
In \cite{MushuUAVTSN},  UAV was leveraged to optimize computation offloading with minimum energy consumption in mobile edge computing.

In terms of the relay inter-satellite links, a routing and wavelength assignment algorithm was proposed to reduce the system cost of inter-satellite laser communication in \cite{XueRoutingTSN}. This work focused on the allocation of optical wavelengths, without considering the optimal minimum number of lasers.
In \cite{BinRoutingTSN}, joint wavelength allocation and routing to decrease the system energy consumption was investigated.
In \cite{PrabhatMaxTSN}, a hybrid satellite-terrestrial relay framework was proposed for a TSN with multi-antenna satellites, and a user-relay selection method was developed to minimize system outage probability.
However, the relationship between the number of lasers for inter-satellite relaying and the resulting relay time needs further investigation.

Motivated by the above observations, we aim to maximize the system energy efficiency in a TSN while considering: 1) relays of hot air balloons between satellites and ground stations at {\color{blue}different heights and minimum elevation angles}; 2) the relation between the delay of inter-satellite relay and the number of configured lasers per satellite; 3) integrated energy efficiency considering the 3C in the integrated TSN.

{\color{blue}We first model the system, in which hot air balloons hover at different heights with different elevation angles.}
The time window between a hot air balloon and a satellite is determined by the minimum elevation angle and their distance. Thus, the time windows are heterogeneous among satellites due to the different heights of hot air balloons.
For inter-satellite links, as there are multiple source, relay, and target satellites, the amount of traffic to be relayed among satellites can be represented by a traffic matrix.
{\color{blue}
%Define the satellite transmitting data traffic  as the source satellite, and the satellite receiving the data traffic  as the target satellite.
We propose an algorithm to determine the traffic matrix from finding multiple sub-traffic-matrices (STMs) in a water-filling manner \cite{LianZhao, LianZhaoGrid}.
The allocation guarantees that the relaying happens during the time windows of the involved source and target satellites.
The next step is to relay the data within STMs to the corresponding target satellites. This step is completed by a well-designed configuration matrices generation algorithm.
We derive the relationship between the relay time and the number of lasers, based on which we can sufficiently utilize the relay time and reduce the number of lasers used for inter-satellite communications.}
Finally, we maximize the system energy efficiency considering constraints including the number of lasers and the traffic delay, \emph{etc}. By geometric programming with Taylor
series approximation, we derive the optimal parameters to maximize the system energy efficiency.

{\color{blue}Our main contributions can be summarized as follows:}
\begin{enumerate}
  \item we propose an algorithm in a water-filling manner \cite{LianZhao, LianZhaoGrid} to relay data among satellites. The proposed algorithm can guarantee that the data is relayed during the time windows of the involved source and target satellites, thereby avoid the overflow.
  \item  we propose a configuration matrices generation method to relay data traffic  in multiple schedules. We also derive the optimal number of lasers subject to the available time for relay among satellites, and achieve a tradeoff between the delay and the required number of lasers for inter-satellite links.
  \item we maximize the system energy efficiency considering various real-world constraints, including the period of a satellite serving a ground station, the transmission power of ground stations,  hot air balloons, and satellites, and the number of lasers at each satellite, \emph{etc}. The novelty of our system model is that the hot air balloons hover at different height with different minimum elevation angle. To the best of our knowledge, this is the first work to address the difference in hovering height and minimum elevation angle problem of hot air balloons.
\end{enumerate}

The remainder of this paper is organized as follows.
Section II presents the network model.
Section III presents the problem formulation.
Section IV presents the STMs determination.
Section V presents the configuration matrices generation method.
Section VI solves the collaborative multi-resource allocation optimization problem.
Numerical results demonstrate the performance gain of the proposed algorithms in Section VII.
We conclude the paper in Section VIII.

\emph{Notation}: Standard notations are used in this paper. $\left| \emph{\textbf{x}} \right|$ is the number of entries in vector $\textbf{\emph{x}}$, and $\textbf{\emph{x}}(\emph{y})$ is the \emph{y}-th element in \textbf{\emph{x}}, where $1 \le y \le \left| \emph{\textbf{x}} \right|$. $\left\lceil  \cdot  \right\rceil $ and $\left\lfloor  \cdot  \right\rfloor $ denote rounding up and rounding down operations on a real number, respectively.

\section{Network Model}

In this section, we introduce the network model of the considered TSN by segments, starting from the hot air balloons.

\subsection{Hot Air Balloons}
\begin{figure}[htbp]
\centering
\begin{minipage}[t]{0.46\textwidth}
\centering
\includegraphics[height=1.8 in]{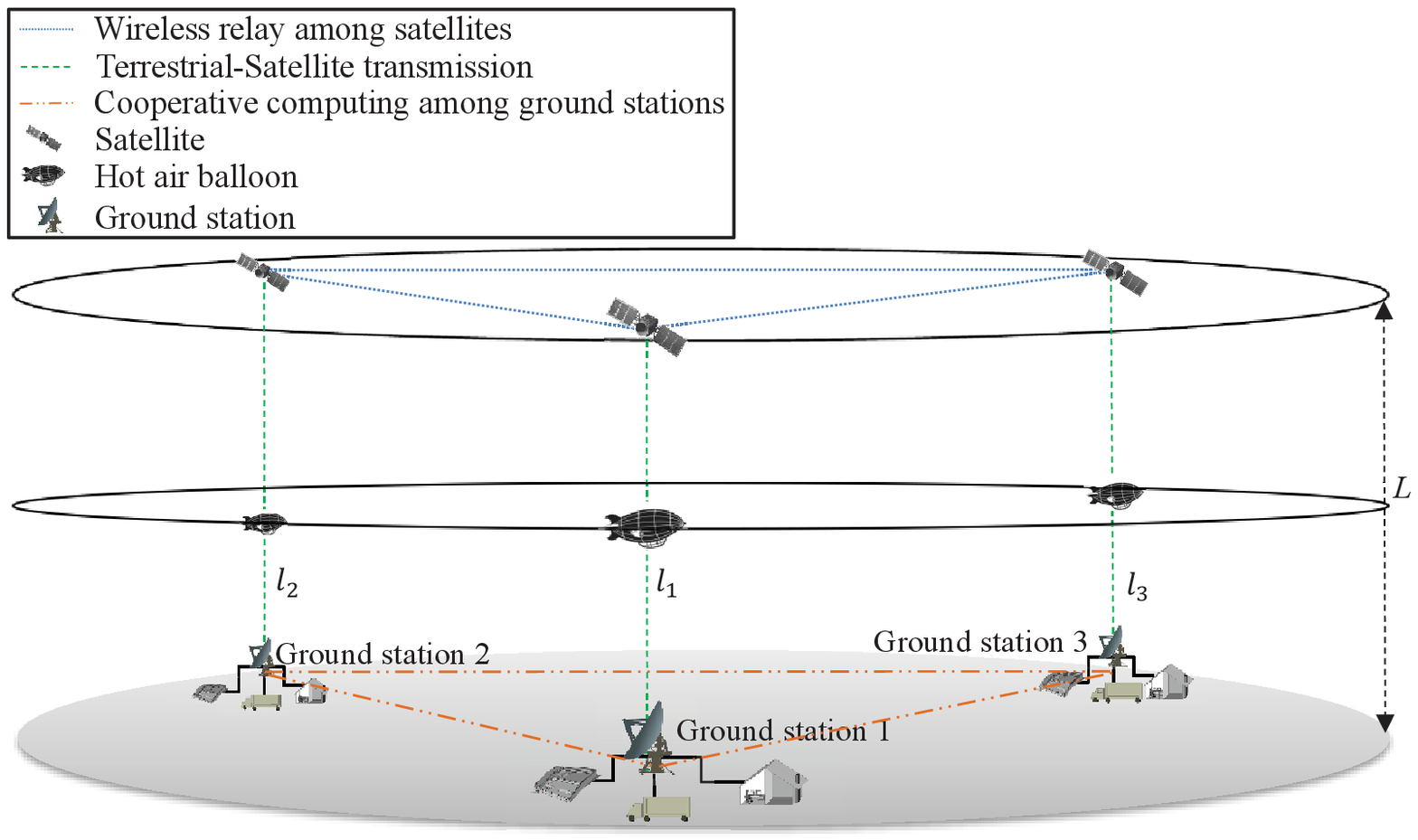}
\renewcommand{\captionlabeldelim}{.}
\centering
\caption{System model.}
\label{fig:System}
\end{minipage}
\begin{minipage}[t]{0.46\textwidth}
\centering
\includegraphics[height=3 in]{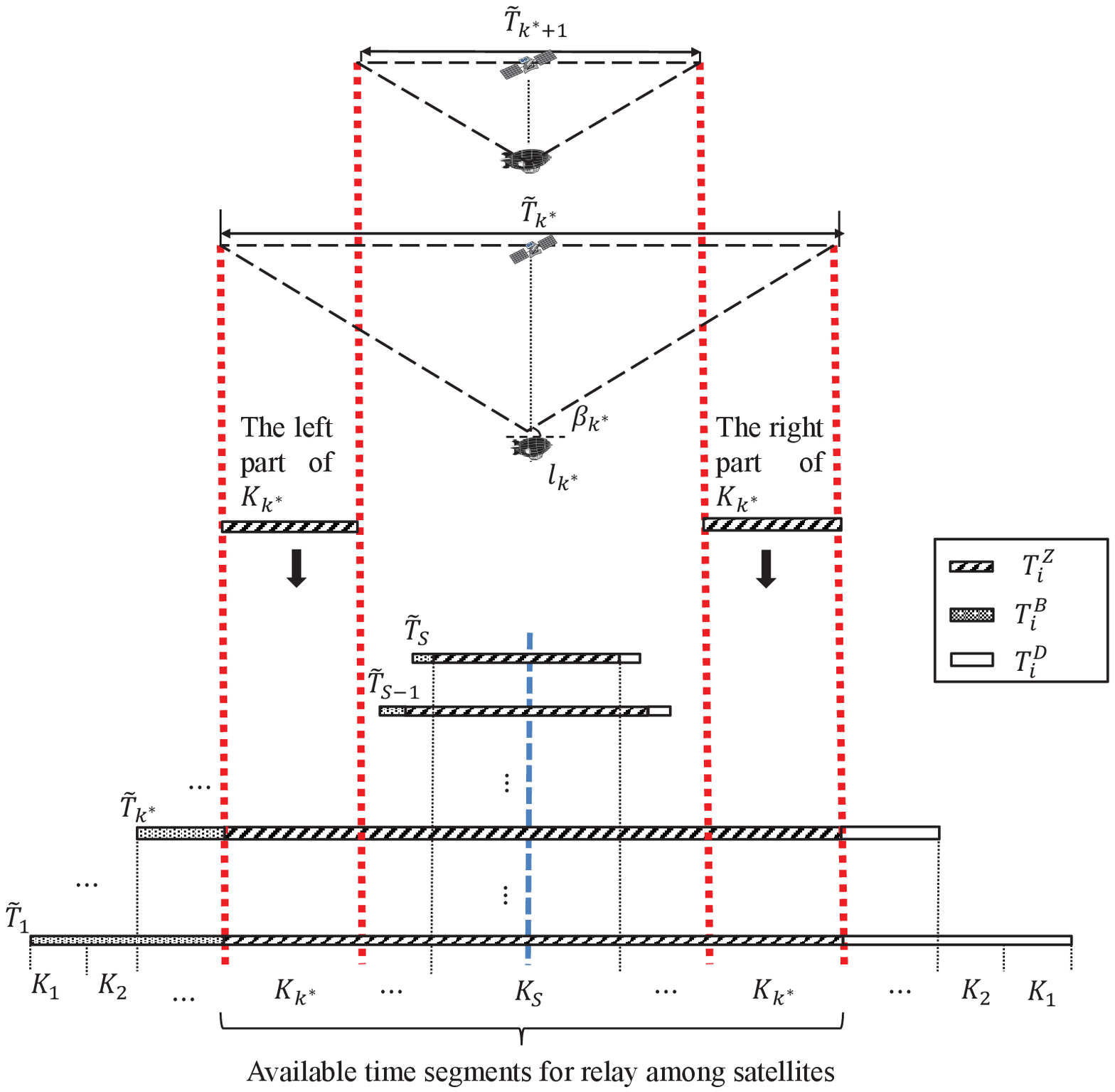}
\renewcommand{\captionlabeldelim}{.}
\centering
\caption{{\color{blue}LEO Satellite Model.}}
\label{fig:Topology}
\end{minipage}
\end{figure}
%\begin{figure}[!tb]
%{\includegraphics[height=2 in]{System.eps}}
%\captionsetup{font={footnotesize }}
%\renewcommand{\captionlabeldelim}{.}
%\centering
%\caption{System model.}
%\label{fig:System}
%\end{figure}
%\begin{figure*}[!tb]
%{\includegraphics[height=3.8 in]{Topology.eps}}
%\captionsetup{font={footnotesize }}
%\renewcommand{\captionlabeldelim}{.}
%\centering
%\caption{LEO Satellite Model.}
%\label{fig:Topology}
%\end{figure*}
The considered system model is illustrated in Fig. \ref{fig:System}, where \emph{S} satellites constitute  a low earth orbit (LEO) satellite network.
 The satellites provide service for \emph{S} ground stations and each satellite has a dedicated ground station.
Denote the number of satellites by \emph{S}.
We assume that the distance from all the satellites to the ground is the same, which is denoted by \emph{L}.
All satellites have the same orbital period given by
\begin{equation}
T^\text{LEO}=2\pi\sqrt{\frac{\left(L+r_E\right)^3}{\mu}},
\end{equation}
where $\mu$=398,601.58 $\text{km}^3/\text{s}^2$ is the Kepler constant, and $r_E$ is the radius of the Earth.
One hot air balloon is hovering above and serving its ground station to play the role of  a dedicated relay between its ground station and the corresponding satellite.
Considering the practical wireless and geographical environment, the distance from the hot air balloons to the ground stations generally differs with each other.
We denote the distance from hot air balloon \emph{i} ($1\leq i\leq S$) to its ground station by $l_i$.
%\begin{equation}
%l_{\text{max}}\leq l_i\leq l_{\text{max}},
% \label{LValue}
%\end{equation}

Due to the orbiting motion of LEO satellites, a hot air balloon can connect with a satellite periodically.
{\color{blue}The continuous serving time between the hot air balloon and the satellite in the $T^\text{LEO}$ is called a time window, which is determined by the minimal elevation angle from the hot air balloon to the satellite.}
An air balloon can exchange data with its satellite only within the corresponding time window.

Specifically, denote the minimal elevation angle from the hot air balloon \emph{i} to its satellite by $\beta_i$ and the geocentric angle of the satellite \emph{i} by $\alpha_i$.
{\color{blue}Generally, a hot air balloon at a lower height has a larger
minimum elevation to the satellite due to the thicker atmosphere for the wireless transmission between the hot air balloon and the satellite.}
Then, it holds that \cite{ADDINGSHU1}:
\begin{equation}
\alpha_i=\arccos\left(\frac{r_E+l_i}{L+r_E}\cos \beta_i\right)-\beta_i.
\end{equation}
The time window of the satellite \emph{i} denoted by $\widetilde{T}_i$, where $\widetilde{T}_i\leq T^\text{LEO}$, is given by
\begin{equation}
\widetilde{T}_i=\frac{2\alpha_i}{2\pi}T^\text{LEO}
=2\left(\arccos\left(\frac{r_E+l_i}{L+r_E}\cos \beta_i\right)-\beta_i\right)\sqrt{\frac{\left(L+r_E\right)^3}{\mu}}.
\label{ConT}
\end{equation}

%As the observation in eq. (\ref{ConT}), the larger $l_i$ leads to the smaller  $\widetilde{T}_i$.
%This suggests that a tradeoff exists between $\widetilde{T}_i$ and $l_i$ under the constraint of the system power.
%On one hand, the smaller $l_i$ leads to the smaller transmission power from the ground stations to the hot air balloon.
%Besides, the larger $\widetilde{T}_i$ can be obtained to provide the sufficient time for traffic relay among the LEO satellite network.
%This effectively decrease the power consumption and system cost involved in the traffic relay.
%On the other hand, the larger $l_i$ pushes the hot air balloon approaching the satellite during its time window. This can effectively decrease the transmission time between the satellite and the hot air balloon, as well as the transmission power.

%satellites and hot air balloons are configured with computing servers. As shown in Fig. \ref{fig:System}, satellites can connect with each other via a dedicated channel of satellites to constitute a cooperative computing resource pool. This can provision the computing ability to keep the relative distance among satellites against orbit perturbation, \emph{etc.}, by adjusting the orbits of satellites in real-time.
The ground stations are connected through a wired network to form a cooperative computing resource pool as shown in Fig. \ref{fig:System}.
This network has the computing ability to generate and execute the scheme of collaborative multi-resource allocation to be developed.
\subsection{LEO Satellites}
The LEO satellite network considered in this work is shown in Fig. \ref{fig:Topology}.
We assume that, when  the satellites circle the earth, they simultaneously arrive at the positions right above their corresponding hot air balloons.
This is due to the identical orbital period $T^{\text{LEO}}$ for all satellites.
As shown in Fig. \ref{fig:Topology}, the time window for each satellite is equally divided into two parts by the straight line from the ground station to the hot air balloon.

Assume that the time windows of the \emph{S} satellites in a LEO satellite network have $\widetilde{T}_1\geq\widetilde{T}_2\geq\ldots\geq\widetilde{T}_S$.
{\color{blue}As shown in the lower part of Fig. \ref{fig:Topology}, the time windows for satellites are symmetrical with respect to the vertical dashed line in the center, which denotes the instant when the satellites are right above their ground stations and hot air balloons.}
Within the time window of satellite $i$, the following tasks should be completed: uplink transmission from the $i$th hot air balloon to the $i$th satellite, which takes a time duration of $T^B_i$; relay for the $i$th satellite, which takes a time duration of $T^Z_i$; and downlink transmission from the $i$th satellite to the $i$th hot air balloon, which takes a time duration of $T^D_i$.
As shown in the lower part of Fig. \ref{fig:Topology}, with $\widetilde{T}_1\geq\widetilde{T}_2\geq\ldots\geq\widetilde{T}_S$, we assume that $T^Z_1\geq T^Z_2\geq\ldots T^Z_S$.

{\color{blue}As shown in the bottom part of Fig. \ref{fig:Topology}, as there are  \emph{S} satellites and \emph{S} time windows, the largest time window can be divided into \emph{S} time segments.
The time segments with sequence numbers $\{k^*, k^*+1, \ldots, S\}$ are employed for traffic relay among satellites, where the value of $k^*$ can be determined by an iterative manner according to the system performance.
Denote the $v$th time segment by $K_v$.
Time segment $K_v$, $v\in\{k^*, k^*+1, \ldots, S-1\}$,  is equally divided into two parts, which are symmetrical with respect to the central dashed
straight line. In the sequel, the time length of time segment $K_v$ refers to the total length of these two parts, denoted as $w_v$.}

For the wireless channel between a satellite and a hot air balloon, as well as that between a hot air balloon and a ground station, we use the omnidirectional path loss model as in existing works \cite{ADDINGSHU1, SBand1, SBand2, SBand3}. The path-loss $\mathcal{C}$ (dB) for a  link with distance \emph{l} is denoted as:
\begin{equation}
\mathcal{C}=92.44+20\times\log_{10}l+20\times\log_{10}f,
\label{PATHLOSS}
\end{equation}
where \emph{f} is the system operating frequency and equals to 3 GHz in the S-band. The unit of the distance \emph{l} is kilometer (km), and the unit of the frequency \emph{f} is GHz in the above path loss model.
Due to the line of sight (LOS) channels between the ground stations and the hot air balloons, and that between the hot air balloons and the satellites, we consider only path loss in the wireless channel power gain, $|H|^2$, as follows.
\begin{equation}
|H|^2=10^{-\frac{\mathcal{C}}{10}}.
\label{LOSCHANNEL}
\end{equation}

For the LEO satellite network, we use the Starlink project launched by SpaceX as a reference \cite{Starlink}.
LEO satellites use inter-satellite-links (ISLs) to form a satellite backbone network in a lattice grid manner \cite{LinCai}.
%Starlink constellation has 1,584 satellites in 72 orbital planes, and 22 satellites in each plane. The orbit is 550 km above earth \emph{i.e.}, \emph{L}=550 km.
For the inter-satellite links, we assume that Terahertz signal \cite{Terahertz1, Terahertz2, Terahertz3} or optical signal is employed.
Each satellite is configured with multiple lasers to generate the signals at high frequency, and multiple antennas to transmit and receive the signals.

\begin{figure}[!htbp]
{\includegraphics[height=3 in]{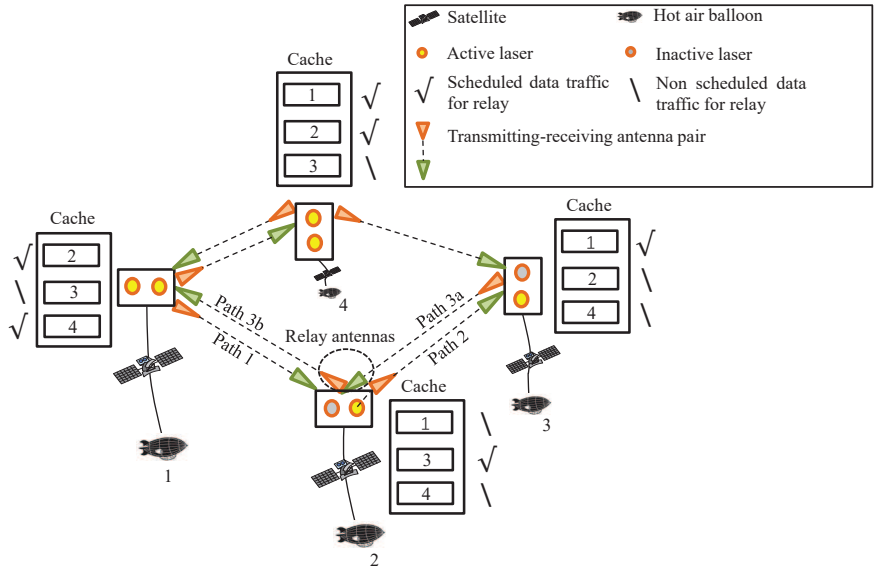}}
\renewcommand{\captionlabeldelim}{.}
\centering
\caption{{\color{blue}Relay among satellites.}}
\label{fig:SatelliteRelay}
\end{figure}

Signals working at high frequency are highly directional due to the narrow wave beam.
As an instance, in Fig. \ref{fig:SatelliteRelay}, the transmitting-receiving antenna pair must be aligned.
This feature can lead to the requirement of multiple schedules  to relay data, because of the limited number of lasers  and antenna pairs at each satellite.
{\color{blue}In Fig. \ref{fig:SatelliteRelay}, each satellite has a cache for storing the data to be transmitted to other satellites.
We call a laser currently generating a high frequency signal as an active laser, and a laser that are currently not in use as an inactive laser.}
In the example shown in Fig. \ref{fig:SatelliteRelay}, let the maximum number of lasers be 2. Then, at least two schedules are needed to complete a full traffic relay cycle for this four satellites network.
 For example at satellite 1, the two lasers can generate the signals to be transmitted to satellite 2 and satellite 4, respectively. Thus, the data traffic  heading to the satellite 3 will be transmitted in the next schedule.

%\begin{figure}[!tb]
%{\includegraphics[height=1.5 in]{Starlink.eps}}
%\captionsetup{font={footnotesize }}
%\renewcommand{\captionlabeldelim}{.}
%\centering
%\caption{LEO network of Starlink project}
%\label{fig:Starlink}
%\end{figure}
%\begin{figure}[!htbp]
%{\includegraphics[height=2.6 in]{SatelliteRelay.eps}}
%\captionsetup{font={footnotesize }}
%\renewcommand{\captionlabeldelim}{.}
%\centering
%\caption{Relay among satellites.}
%\label{fig:SatelliteRelay}
%\end{figure}
\subsection{Traffic Relaying Among Satellites}
Traffic relaying among satellites can be illustrated as shown in Fig. \ref{fig:SatelliteRelay}.
Each data traffic stream is generated by a laser and sent by a transmitting antenna. The data traffic  is relayed among satellites and received by the receiving antenna of the target satellite.
In practice, one hop of the transmission can be built between two satellites with the distance less than or equal to a pre-determined threshold.
If the distance from the source satellite to the target satellite is larger than the threshold, a relay satellite is required.
For instance, in Fig. \ref{fig:SatelliteRelay}, satellite 3 sends its data to satellite 1 via path 3 consisting of path 3a and path 3b with satellite 2 as the relay.
%In Fig. \ref{fig:SatelliteRelay}, the average active number of lasers per satellite is $\overline{m}=2/3$, and the average active number of transmitting-receiving pairs per satellite is $\overline{w}=3/3=1$.
The inter-satellite data traffic is represented by a matrix $\textbf{\emph{A}}=\{a_{ij}\}$, where $a_{ij}$ is the amount of data from the source satellite \emph{i} to the target satellite \emph{j}.
We have $a_{ii}=0$ because satellite \emph{i} does not need to send data to itself.

We assume that the number of schedules of the data is $\vartheta$.
Denote the average active number of lasers per satellite by $\overline{m}$, which is bounded by the maximum value of $m_{\text{max}}$.
Denote the power to launch a laser by $P^Z$.
Denote the delay of aligning antennas before a scheduled transmission by $\delta_Y$.
The basic energy consumption including circuit energy consumption, \emph{etc.}, for lasers of the satellites  is
\begin{equation}
\varepsilon^Z=P^ZS\overline{m}\delta_Y\vartheta.
\label{EEZ}
\end{equation}
%In this paper, we assume that
%\begin{equation}
%\overline{m}\leq m_{\text{max}}.
%\label{mMAX}
%\end{equation}

The velocity of electromagnetic signal propagation in space is $V = 3\times10^8$ m/s.
Define the maximal distance among all paths of source-target satellites by $\Omega$. Then,  the overhead involved in data transmission among satellites is $\delta_R=\Omega/V$.
Prior to a scheduled transmission, overhead $\delta$ exists, which includes the alignment of antennas among satellites $\delta_Y$, and the data transmission time on paths of source-target satellites $\delta_R$. The total overhead time, denoted by $\delta$, is formulated as:
\begin{equation}
\delta=\delta_Y+\delta_R.
\label{Overhead}
\end{equation}
In eq. (\ref{Overhead}), for simplification, we assume that $\delta_Y$ is the same for each satellite.
Since we consider the upper bound of $\delta_R$, $\delta$ can be treated as a constant in each schedule.

In Fig. \ref{fig:SatelliteRelay}, a pair of transmitting-receiving antennas can act as the relay for data transmission.
The average traffic relay rate can be calculated by $\sum_{i=1}^S\sum_{j=1}^Sa_{ij}/T^Z_{k^*}$, where $T^Z_{k^*}$ is the maximum relay time as shown in Fig. \ref{fig:Topology}.
Denote the static power consumption of one laser by $P_0^L$ W/bps, then, the static energy consumption of lasers is
\begin{equation}
\varepsilon^L_0=\left(\sum_{i=1}^S\sum_{j=1}^Sa_{ij}/T^Z_{k^*}\right)\times\overline{m}P_0^LS\times\widetilde{T}_{k^*}.
\label{StaticLaser}
\end{equation}
Denote the dynamic power consumption of one laser by $P_1^L$ W/bps, and the capacity of the inter-satellite channels by $C_0$ bps.
Then, the dynamic energy consumption of lasers is
\begin{equation}
\varepsilon^L_1=C_0\times\overline{m}P_1^LS\times T^Z_{k^*}.
\label{DynamicLaser}
\end{equation}
In practice, $P_1^L$ may depend on the distance between satellites and the signal frequency. In this work, $P_1^L$ is assumed to be a constant for the considered TSN.

\section{Problem Formulation}

In this section, we present the data process steps involved in the considered TSN and formulate a network energy efficiency maximization problem considering  constraints on 3C resources.
\subsection{Data Transmission and Relay}
A service period in the system consists of 6 steps and is described as follows.
\begin{enumerate}[Step 1)]
  \item Accumulation of data at ground stations. Data is gathered and accumulated at the ground stations for time $T^A$ before each transmission, with $T^A$ given by
      \begin{equation}
      T^A=n_0\times T^{\text{LEO}},
      \label{TAi}
      \end{equation}
      where $n_0\geq1$ represents the number of circles the satellite orbiting over the Earth in between two transmissions, and $n_0 T^{\text{LEO}}$ is referred to as the serving period of satellites.
      We denote the maximum value of  $n_0$ by $n_{\text{max}}$ so that $n_{\text{max}}T^{\text{LEO}}$ is the maximum tolerable system delay at the ground stations.
      Denote the average data arrival rate by $\lambda_i$ and the power consumption for caching data during the accumulation duration by $P^A$ W/bit.
      Then, the energy consumption for caching the data at the \emph{i}th ground station is  $P^A\lambda_iT^A$.
      %Define $a_{ij}$ as the amount of traffic to relay from satellite \emph{i} to satellite \emph{j} in one LEO satellite orbital period \emph{i.e.}, $n_0=1$.
%      The corresponding average traffic matrix for $n_0=1$ is $\textbf{\emph{A}}=\{a_{ij}\}$.
%      Then, the average traffic matrix accumulated in $n_0T^{\text{LEO}}$  ($n_0\geq1$) can be formulated by $\textbf{\emph{A}}'=\{a'_{ij}\}=n_0\textbf{\emph{A}}$.
  \item Calculating the traffic scheduling and routing scheme.
   After the traffic accumulation, the computing servers of the ground stations  cooperatively calculate the scheme of traffic scheduling and routing with a pre-determined algorithm.
   {\color{blue}Denote the computing capacity of the cooperative computing pool at the ground stations by $C^{\text{HAB}}$ cycles/second (cps) and the  computing demand of the data scheduling and routing in a network with \emph{S} satellite by $\eta S$ cycles, where $\eta$ is the average computing load for adding one satellite in the satellite relay network.} The computing delay $T^C$ is give by
       \begin{equation}
      T^C=\frac{\eta S}{C^{\text{HAB}}}.
      \label{ComputingDelay}
      \end{equation}
  In terms of the energy consumption of the computing resource, we denote the power consumption of the computing by $P^C$ W/cps. The corresponding energy consumption of computing $\varepsilon^C$ is
    \begin{equation}
     \varepsilon^C=P^C\eta S.
      \label{ComputingEnergy}
      \end{equation}
  \item Transmission of data traffic  from  ground stations to hot air balloons. {\color{blue}Let $T^G_i$ denote the data transmission time from the $i$th ground station to its serving hot air balloon with transmission power $P^G_i$.} Assume that the transmission bandwidth is $B_0$, \textcolor{blue}{which is identical for all ground station to hot air balloon transmissions}, and the power spectral density of the Gaussian white noise is $\sigma^2$ W/Hz.
      By eq. (\ref{LOSCHANNEL}), the wireless channel power gain is denoted by $|H^G_i|^2=10^{-\frac{106.3+20\times\log_{10}l_i}{10}}$.
      Denote the antenna gain by $G_T$, with $|H^G_i|^2$, we can prove that
       \begin{equation}
       P^G_i=\frac{B_0\sigma^2l_i^{2}\times10^{11.44}\left(2^{\frac{n_0T^{\text{LEO}}\lambda_i}{B_0T^G_i}}-1\right)}{G_T}.     \label{PTG}
       \end{equation}
       The delay caused by the transmission is $t_1=l_i/V$.
       In order to efficiently utilize the time window of $\widetilde{T}_i$ for the inter-satellite relay and the rely between hot air balloons and satellites, we have following constraint,
      \begin{equation}
      T^G_i+T^C+\frac{l_i}{V}=
      T^G_i+\frac{\eta S}{C^{\text{HAB}}}+\frac{l_i}{V}\leq T^{\text{LEO}}-\widetilde{T}_i, \forall i.
      \label{TBTQ}
      \end{equation}
      Constraint (\ref{TBTQ}) means that the delay in step 2) plus step 3) should be within $T^{\text{LEO}}-\widetilde{T}_i$.
      %This constraint allows efficiently utilizing the time windows for the data exchange between hot air balloons and satellites, as well as relay among satellites.
      Accordingly, the following three steps should be completed within the corresponding time windows.
  \item Data transmission from hot air balloons to satellites.
     The transmission time of the \emph{i}th hot air balloon to the corresponding serving satellite is denoted by $T^B_i$, which should be a portion of the time window of satellite $i$.
      Denote the transmission power of the \emph{i}th hot air balloon by $P^B_i$.  Denote the transmission bandwidth of a hot air balloon by $B_1$, which is identical for all balloon to satellite transmissions.
      Denote the path loss based wireless channel power gain by $|H^B_i|^2=10^{-\frac{106.3+20\times\log_{10}\left(L-l_i\right)}{10}}$. Then, with $G_T$ and $|H^B_i|^2$, we can prove that
       \begin{equation}
       P^B_i=\frac{B_1\sigma^2\left(L-l_i\right)^{2}\times10^{11.44}\left(2^{\frac{n_0T^{\text{LEO}}\lambda_i}{B_1T^B_i}}-1\right)}{G_T}.     \label{PTB}
       \end{equation}
%      \begin{equation}
%      \begin{split}
%      &T^B_i=\\
%      &\frac{n_0T^{\text{LEO}}\lambda_i}{B_1\log_2\left(1+\frac{P^B_i\times10^{-\frac{106.3+20\times\log_{10}\left(L-l_i\right)}{10}}}{\sigma^2B_1}\right)}, \forall i.
%      \label{DelayTB}
%      \end{split}
%      \end{equation}
      The corresponding transmission delay is $t_2=\left(L-l_i\right)/V$.
     % Then, multiple traffic packets at mmwave frequency are formed as one packet at Terahertz frequency according to the source and aimed satellites to sufficiently explore the large capacity of Terahertz wireless channel.
  \item Data relaying among satellites. After the data transmission from hot air balloons to satellites, the inter-satellite relaying is planned in multiple schedules for forwarding data to target satellites.
  \item Data transmission from satellites to hot air balloons.
      After the inter-satellite relaying, the data will be transmitted from satellites to hot air balloons, which takes a duration of length $T^D_i$, and then forwarded to the target ground stations to complete a ground station to ground station data communication cycle.
      Denote the average data traffic arrival rate at satellite \emph{i}, \textcolor{blue}{as a result of inter-satellite relaying}, by $\mu_i$.
      It holds that $\sum_{i=1}^S\lambda_i=\sum_{i=1}^S\mu_i$.
       The amount of the data received at satellite \emph{i} in duration $\widetilde{T}_i$ is then $n_0T^{\text{LEO}}\mu_i$.
      Denote the transmission power of sending data from satellite \emph{i} to its hot air balloon by $P^D_i$. Denote the bandwidth by $B_2$, \textcolor{blue}{which is identical for all satellite to balloon transmissions}.
      The corresponding channel gain is $|H^D_i|^2=10^{-\frac{106.3+20\times\log_{10}\left(L-l_i\right)}{10}}$. Then, with $G_T$ and $|H^B_i|^2$, we have
       \begin{equation}
       P^D_i=\frac{B_2\sigma^2\left(L-l_i\right)^{2}\times10^{11.44}\left(2^{\frac{n_0T^{\text{LEO}}\mu_i}{B_2T^D_i}}-1\right)}{G_T}.     \label{PTD}
       \end{equation}
%      \begin{equation}
%      \begin{split}
%      &T^D_i=\\
%      &\frac{n_0T^{\text{LEO}}\mu_i}{B_2\log_2\left(1+\frac{P^D_i\times10^{-\frac{106.3+20\times\log_{10}\left(L-l_i\right)}{10}}}{\sigma^2B_2}\right)}.
%      \end{split}
%      \end{equation}
      The propagation delay in this step is $t_3=\left(L-l_i\right)/V$.
       \end{enumerate}

\textcolor{blue} {Next, the data collected at  hot air balloons  from satellites will be used for specific objectives such as store-and-forward for hot traffic or transmission to the ground stations for real-time traffic.}

\subsection{Optimization Model}
\begin{figure}[!htbp]
{\includegraphics[height=1.8 in]{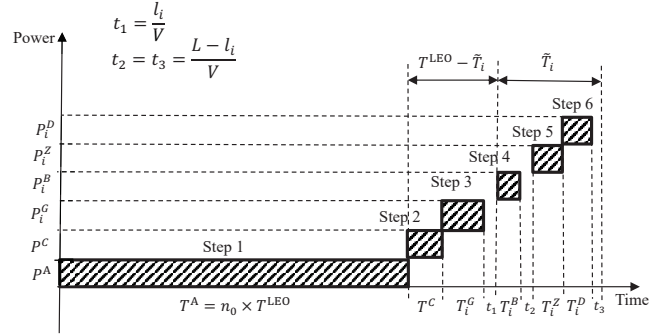}}
\renewcommand{\captionlabeldelim}{.}
\centering
\caption{Illustration of the six steps of a communication cycle.}
\label{fig:Delay}
\end{figure}
In this sub-section, we focus on formulating the system energy efficiency maximization problem.
First, we illustrate the constraints of data transmission in the aforementioned 6 steps as in Fig. \ref{fig:Delay}.The ordinate in Fig. \ref{fig:Delay} is the average power consumption for each step, and the abscissa is the time for each step. The \textcolor{blue}{propagation delay} is denoted by $t_1$ for step 3, $t_2$ for step 4, and $t_3$ for step 6, respectively, as shown in Fig. \ref{fig:Delay}.
To avoid data overflow, $T^A$ in step 1 is the upper bound for each of the next steps.
Since $T^A\geq T^{\text{LEO}}$ as shown in eq. (\ref{TAi}), we have
\begin{equation}
T^C+T^G_i+\frac{l_i}{V}\leq T^{\text{LEO}}-\widetilde{T}_i< T^A,
 \label{XStep23}
\end{equation}
and
\begin{equation}
T^B_i+T^D_i+2\times\frac{L-l_i}{V}+T^Z_i\leq\widetilde{T}_i<T^A.
 \label{XStep456}
\end{equation}
There exists a trade-off between the \textcolor{blue}{delay} and power consumption in each step, which will be discussed later.

%Since the wireless interference among flight equipment is ignored, by eqs. (\ref{XStep23})-(\ref{XStep456}), we can propose Theorem 1 as follows:

%\underline{\textbf{Theorem 1}} When the wireless interference among flight equipment can be ignored, the formulations can be verified to minimize the system energy consumption:
%\begin{equation}
%T^G_i+T^C= T^{\text{LEO}}-\widetilde{T}_i-\frac{l_i}{V}.
% \label{Step23}
%\end{equation}
%\begin{equation}
%T^B_i+T^Z_i+T^D_i= \widetilde{T}_i-2\times\frac{L-l_i}{V}.
% \label{Step456}
%\end{equation}
%
%\emph{Proof}: See Appendix A.
%\hfill\rule{4pt}{8pt}

%Basically, a larger time consumed in one step can lead to a less power consumption as well as the resource consumption.
%The time for each step has an upper bound $T^A$ as shown in Fig. \ref{fig:Delay}.

\begin{figure*}[!b]
\hrulefill
\begin{equation}
\frac{1}{E^{\text{Total}}}=\frac{\sum_{i=1}^S\left(P^A \lambda_i n_0T^{\text{LEO}}+P^G_iT^G_i+P^B_iT^B_i+P^D_iT^D_i\right)
+\varepsilon^C+\varepsilon^Z+\varepsilon^L_0+\varepsilon^L_1}{n_0T^{\text{LEO}}\times\sum_{i=1}^S\lambda_i}.
\label{TotalEnergy}
\end{equation}
\begin{equation}
\emph{\textbf{Var}}=\{n_0, \overline{m}, P^G_i, P^B_i, P^D_i, T^G_i, T^B_i, T^D_i, T^Z_i, \vartheta\}.
\label{Var}
\end{equation}
\hrulefill
\end{figure*}

Denote the  total energy efficiency from Step 1 to Step 6 by $E^{\text{Total}}$.
Considering the analysis above, $1/E^{\text{Total}}$ can be formulated by eq. (\ref{TotalEnergy}).
In eq. (\ref{TotalEnergy}), the denominator is \textcolor{blue}{the total amount of the data  transmitted in one serving period}.
The overall energy consumption, used for caching, computing, wireless transmission (including the lasers),  is formulated in the numerator of eq. (\ref{TotalEnergy}).
As there are many variables, we define a set of all variables in eq. (\ref{Var}).

With eqs. (\ref{TotalEnergy}) and  (\ref{Var}), the target optimization problem can be formulated as
\begin{equation} \tag{P1}
  \begin{split}
  &\mathop{\text{minimize}}\limits_{\emph{\textbf{Var}}}\frac{1}{E^{\text{Total}}},\\
  &\emph{s.t.} \quad (\ref{XStep23})  \quad\text{and}\quad (\ref{XStep456}).
  \end{split}
  \label{P1}
\end{equation}

{\color{blue}Generally speaking, the energy consumption of antenna pairs depends on routing paths. The energy consumption of traffic scheduling for relay among satellites depends on the number of used lasers per satellite.}
In this work, we do not include the energy consumption for routing among the satellites in the free space and the antenna pairs, since this part of energy consumption is relatively small.

\section{{\color{blue}STMs Determination}}
\begin{figure*}[!htbp]
{\includegraphics[height=3.4 in]{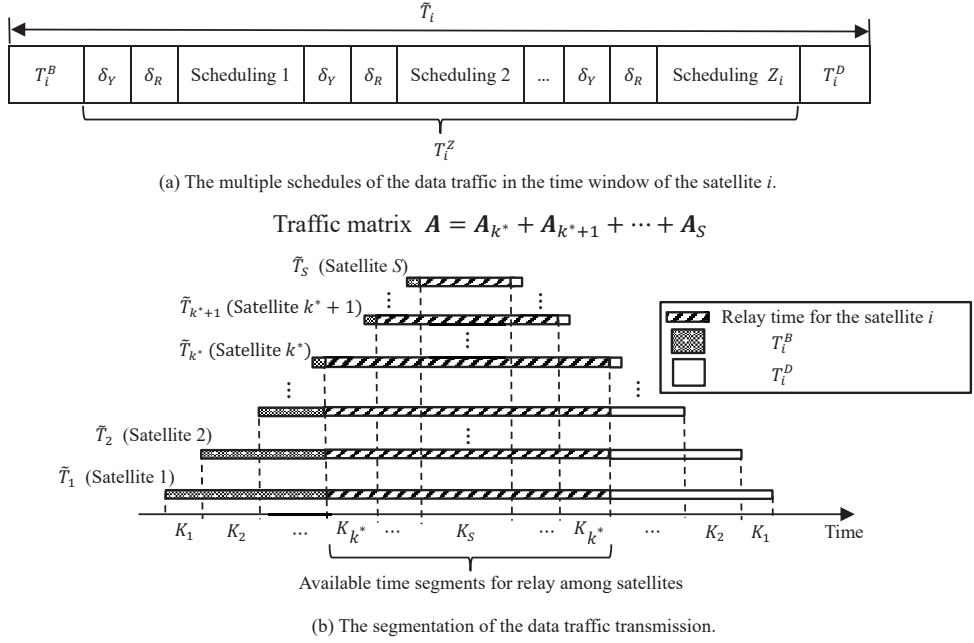}}
\renewcommand{\captionlabeldelim}{.}
\centering
\caption{Traffic segmentation for traffic scheduling.}
\label{fig:SchedulingandRouting}
\end{figure*}
As aforementioned, data relayed via inter-satellite communications constitutes a traffic matrix \textbf{\emph{A}}.
{\color{blue}To solve the optimization problem (\ref{P1}), we first study STMs determination based on different time window lengths of the satellites. The objective of the STMs determination is allocating the data traffic to different time segments. The relay scheduled for each time segment must happen within the time windows of the corresponding source and target satellites.}
Using the case shown in Fig. \ref{fig:SatelliteRelay} for instance, each satellite caches the data to be relayed to the other three satellites.
The different time window lengths of satellite determine the different available times for inter-satellite relaying, and the data for relaying indicated by the traffic matrix should be decomposed into multiple STMs, each for one specific time segment.

As shown in Fig. \ref{fig:SchedulingandRouting} (a), the time window $\widetilde{T}_i$ contains $T^B_i$, $T^D_i$, and relay time $T^Z_i$.
%Each schedule contains the delay overhead $\delta=\delta_Y+\delta_R$.
Without loss of generality, we assume that the length of the time window satisfies
$\widetilde{T}_1\geq\widetilde{T}_2\geq\ldots\geq\widetilde{T}_S$.
%\begin{equation}
%\widetilde{T}_1\geq\widetilde{T}_2\geq\ldots\geq\widetilde{T}_S.
%\label{RET}
%\end{equation}
We also assume that the {\color{blue}maximum available time} for inter-satellite relaying in Step 5 is within the relay time of the $k^*$th satellite, \emph{i.e.}, $T_{k^*}^Z$, which is formulated as
\begin{equation}
T^Z_i=\left\{
\begin{aligned}
\alpha  \widetilde{T}_i& , & i\geq k^*, \\
\alpha \widetilde{T}_{k^*} & , & i<k^*,
\end{aligned}
\right.
\label{TZTZ}
\end{equation}
where $0<\alpha<1$.
By eq. (\ref{TZTZ}), we have $T^Z_1= T^Z_2=\ldots= T^Z_{k^*}$ due to $T^Z_i=\alpha\widetilde{T}_{k^*}$ for $i<k^*$.
When $T^B_i$ and $T^D_i$ are determined, the value of $\alpha$ can be found.
{\color{blue} As we have assumed, the length of the \emph{v}th time segment is $w_v$.
Denote the length of the \emph{v}th time segment by $\tau_v$ when $\alpha=1$.
Then, we have $w_v=\alpha\tau_v$, where $\tau_v$ is a constant determined by the length of time windows, and $\alpha$ will be optimized later.}
%In the following discussion, we let $n_0 = 1$. The result can be extended to general values.

% \underline{\textbf{Definition 1}}
%The basic case of generating the sub-traffic matrices is the traffic matrix with $\alpha=1$ and $n_0=1$.
%\hfill\rule{4pt}{8pt}

The data traffic matrix $\textbf{\emph{A}}$ can be decomposed into \emph{S} STMs $\{\textbf{\emph{A}}_1, \textbf{\emph{A}}_2,\ldots, \textbf{\emph{A}}_S\}$, each representing the data traffic to be relayed in the corresponding time segment.
Since only the time segments with the sequence numbers $\{k^*, k^*+1, \ldots, S\}$ are employed for traffic relay, $\{\textbf{\emph{A}}_1, \textbf{\emph{A}}_2,\ldots, \textbf{\emph{A}}_{k^*-1}\}$ are all-zero matrices, \emph{i.e.}, $\textbf{\emph{A}}=\sum_{i=1}^S\textbf{\emph{A}}_i=\sum_{i=k^*}^S\textbf{\emph{A}}_i$.
The element of the \emph{i}th row and the \emph{j}th column of $\textbf{\emph{A}}_v$ is denoted by $a^v_{ij}$, which represents the amount of data traffic transmitted from the \emph{i}th satellite to the \emph{j}th satellite during time segment $K_v$.
In the rest of this Section, we will solve the STMs generation problem.

\subsection{Geometric-Water-Filling Algorithm}
 \begin{figure}[htbp]
\centering
\begin{minipage}[t]{0.48\textwidth}
\centering
\includegraphics[height=1.6 in]{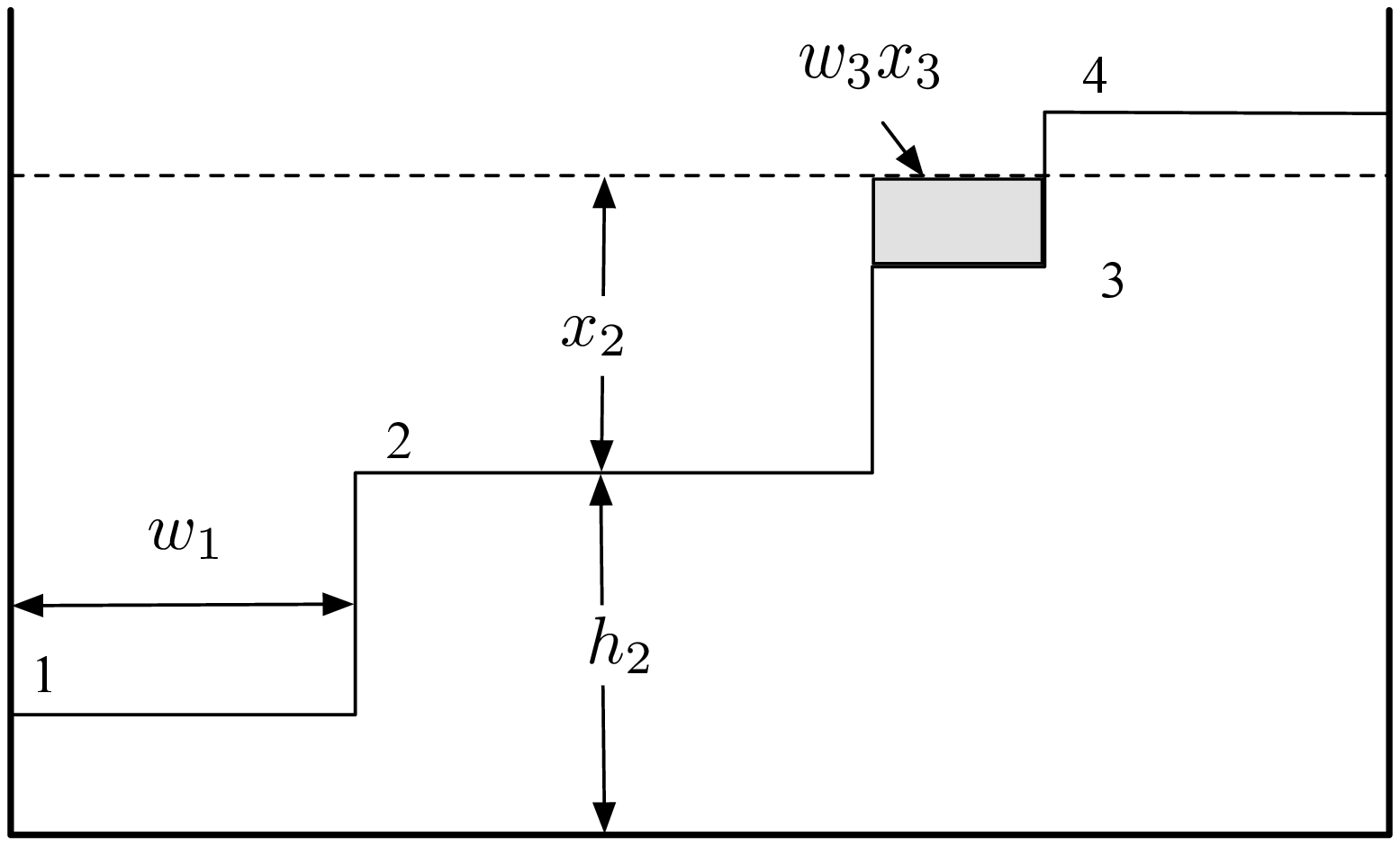}
\renewcommand{\captionlabeldelim}{.}
\centering
\caption{Illustration of GWF algorithm.}
\label{fig:GWF1}
\end{minipage}
\begin{minipage}[t]{0.48\textwidth}
\centering
\includegraphics[height=2.2 in]{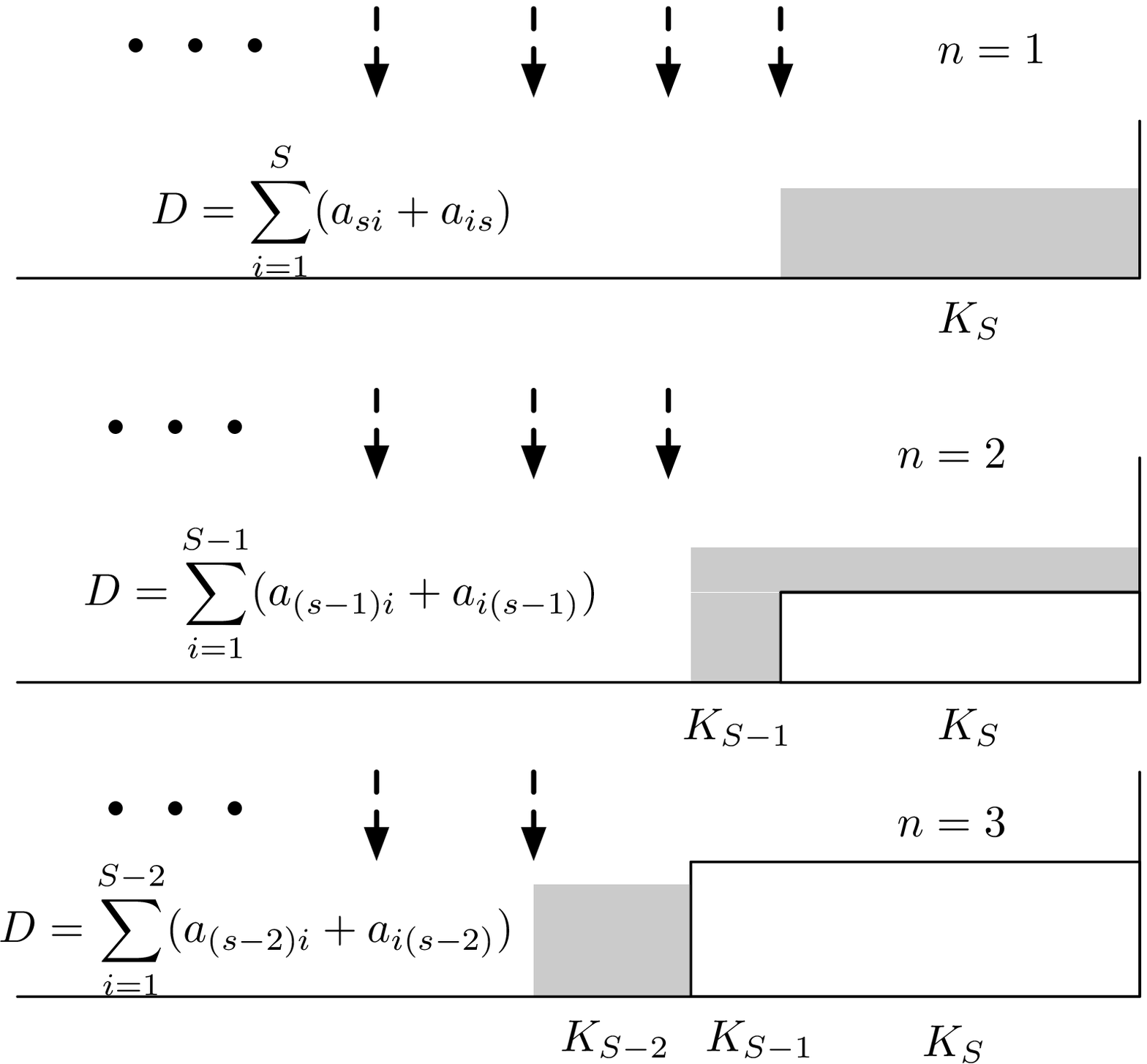}
\renewcommand{\captionlabeldelim}{.}
\centering
\caption{Illustration of T-GWF algorithm.}
\label{fig:T-GWF}
\end{minipage}
\end{figure}
%\begin{figure}[!htbp]
%{\includegraphics[height=1.8in]{GWF1.eps}}
%\captionsetup{font={footnotesize }}
%\renewcommand{\captionlabeldelim}{.}
%\centering
%\caption{Illustration of GWF algorithm.}
%\label{fig:GWF1}
%\end{figure}
%\begin{figure}[!htbp]
%{\includegraphics[height=2.4in]{T-GWF.eps}}
%\captionsetup{font={footnotesize }}
%\renewcommand{\captionlabeldelim}{.}
%\centering
%\caption{Illustration of T-GWF algorithm.}
%\label{fig:T-GWF}
%\end{figure}

{\color{blue}The objective of the STMs determination to different time segments
with various length is to guarantee that the relay time is within the time windows of the corresponding source and destination satellites and to achieve data load balancing.}
Water-filling is well-known for deriving the optimal solutions for power allocation problems \cite{LianZhao} as well as load balancing problems \cite{LianZhaoGrid}. In the following, we introduce the geometric-water-filling (GWF) algorithm. More details can be found in \cite{LianZhao, LianZhaoGrid}.

As shown in Fig.~\ref{fig:GWF1}, we assume a water tank with $k$
uneven bottom steps. The width of the steps is denoted by vector
$ \textbf{\emph{W}}$ and height by vector $ \textbf{\emph{H}}$, representing the time segments width and already allocated data traffic density (\emph{i.e.}, data traffic per unit time), respectively.
The total volume of water, $D$, represents the total amount of data for distribution.
The dashed horizontal line represents the
formed water surface. The optimal distribution of these data
amount is depicted by the water-filling algorithm. The solution returns a
vector, $\textbf{\emph{X}}$, the height of the distributed water above the steps,
denoting the added traffic density for the corresponding time
segment. In Fig.~\ref{fig:GWF1}, the 4th step height $h_4$ is
higher than the water surface, the solution of $x_4$ is then zero.

We can represent the geometric water-filling function as
 \begin{equation}
   \textbf{\emph{X}} =\mbox{GWF}(k,  \textbf{\emph{W}}, \textbf{\emph{H}} , D).
   \label{eqn:GWF1}
 \end{equation}
The data amount scheduled for the $i$th time segment is $w_i x_i$.
The total data amount satisfies $D=\sum_{i=1}^{k} w_i x_i$. The
detailed solution approach can be referred in \cite{LianZhao, LianZhaoGrid}.
Without loss of focus, we shall directly apply eq.
(\ref{eqn:GWF1}) as a functional block to solve the STMs for satellites
traffic relay.

\begin{algorithm}[!tb]
\footnotesize
\caption{T-GWF based STMs determination algorithm.}
\label{alg:TWF}
\begin{algorithmic}[1] %这个1 表示每一行都显示数字
\renewcommand{\algorithmicrequire}{\textbf{Initialization and Input:}}
\REQUIRE ~~\\ %算法的输入参数：Input
Initialization:
$k^*$: the index of the widest time window being used for relay;
 $\textbf{\emph{A}}$: traffic matrix $(S \times S)$;
$\textbf{\emph{A}}_v$ : zero matrices $(S \times S)$, $v=k^*, \cdots, S$ (STMs);
 $\emph{\textbf{W}}=[0, \cdots, w_{k^*}, \cdots, w_{S}]$: width of the time segment;\\
 $\emph{\textbf{H}}$ = zeros($1: S$);
 $D_{\mbox{total}} = \sum_{i=1}^{S} \sum_{j=1}^{S} a_{ij}$; \\
 $\emph{\textbf{D}}_{\mbox{allocated}} = 0$;
$n=1$.
\FOR{($m=S: -1: k^*$)}
\IF{$m>k^*$}
\STATE
$\emph{\textbf{D}}_{\mbox{row}} =\textbf{\emph{A}}(m,1:m) $;
\STATE
$\emph{\textbf{D}}_{\mbox{col}} =\textbf{\emph{A}}(1:m,m) $;
\STATE
$D= \sum_{i=1}^m(\emph{\textbf{D}}_{\mbox{row}}(i)+\emph{\textbf{D}}_{\mbox{col}}(i))$;
\STATE
$\emph{\textbf{D}}_{\mbox{allocated}} \leftarrow D_{\mbox{allocated}} +D$;
\ELSE
\STATE
$D=D_{\mbox{total}} - D_{\mbox{allocated}}$;
\STATE
 $\emph{\textbf{D}}_{\mbox{matrix}} = \emph{\textbf{A}}(1:k^*, 1:k^*)$;
\ENDIF
\STATE
 $\emph{\textbf{X}}= \mbox{GWF} (n, \emph{\textbf{W}}(m:S), \emph{\textbf{H}}(m:S), D)$;
\STATE
$\emph{\textbf{H}}(m:S) \leftarrow \emph{\textbf{H}}(m:S) +\emph{\textbf{X}}$;
\STATE
$n \leftarrow n+1$;
\FOR{($v=m:S$)}
\STATE
$\eta =W(v)\cdot X(v)/D$;
\IF{$m>k^*$}
\STATE
$\textbf{\emph{A}}_v(m,1:m) \leftarrow\textbf{\emph{A}}_v(m,1:m) +  \widetilde{\eta} \cdot\emph{\textbf{D}}_{\mbox{row}}$;
\STATE
$\textbf{\emph{A}}_v(1:m, m) \leftarrow\textbf{\emph{A}}_v(1:m, m) + \widetilde{\eta} \cdot\emph{\textbf{D}}_{\mbox{col}}$;
\ELSE
\STATE
$\textbf{\emph{A}}_v(1:k^*, 1:k^*) \leftarrow\textbf{\emph{A}}_v(1:k^*, 1:k^*) + \widetilde{\eta} \cdot \emph{\textbf{D}}_{\mbox{matrix}}$;
\ENDIF
\ENDFOR
\ENDFOR
\STATE \textbf{Return}: $\textbf{\emph{A}}_v$, $k^*\leq v\leq S$. % 算法的返回值
\end{algorithmic}
\end{algorithm}

\subsection{Tapped Geometric-Water-Filling Based STMs Determination}
In this subsection, we present the proposed algorithm for
STMs determination. During the time window for
the $k^*$th satellite, we have a series of time segments, denoted
as $K_{k^*}, \cdots, K_S$ with width $w_{k^*}, \cdots, w_S$
respectively. Data of the total traffic matrix $\textbf{\emph{A}}$ is
allocated to STMs, $\textbf{\emph{A}}_{k^*}, \cdots, \textbf{\emph{A}}_S$, one for each time segment.
%In order to decrease the required number of lasers, the traffic
%data should be balanced among different time segments.
We propose tapped geometric-water-filling (T-GWF) algorithm as described  in Algorithm 1 to generate the STMs.

In the initialization stage of Algorithm 1, we set all STMs as all-zero matrices. The elements of the time segment width vector $\textbf{\emph{W}}$ are set to
0 for time segments $K_{1}$ to $K_{k^*-1}$ and the corresponding true width for time segments $K_{k^*}$ to $K_S$.
$D_{\mbox{total}}$ and $D_{\mbox{allocated}}$ represent the total
data amount and the allocated amount, respectively.

In Line 1, a ``For" loop is used, running from the $K_S$th time
segment to the $K_{k^*}$th time segment. Before the $K_{k^*}$th time
segment, we start from time segment $K_S$, where the $S$th row and the
$S$th column of the traffic matrix $\textbf{\emph{A}}$ is stored in
$\emph{\textbf{D}}_{\mbox{row}}$ and $\emph{\textbf{D}}_{\mbox{col}}$ respectively
in Lines 3 and 4. The total amount of data for the allocation in
this round is denoted by $D$ as shown in Line 5.

In Line 11, the GWF algorithm is applied to conduct data traffic
allocation. The notation $\emph{\textbf{W}}(m:S)$ denotes a vector
consisting of  the $m$th element to the $S$th element of vector
$\emph{\textbf{W}}$. We design a mechanism recursively updating the STMs
through GWF operation. We model that the time segments are
separated by a series of taps, shown as the vertical arrowed
dashed-lines in Fig.\ref{fig:T-GWF}. Starting from $n=1$, when GWF
works for the $K_S$th segment, and all the taps are in their closed
positions. Then when $n=2$, the right most tap is removed and the
data traffic  can be distributed in time segments $K_{S-1}$ to
$K_S$. For all the time segments involved in the traffic allocation,
the newly allocated traffic is frozen and added on the top of the
current step height to update the step height for the next round
GWF calculation. The taps are lifted one by one from right to left, until the last tap for the time segment $K_{k^*}$. We
refer to the proposed Algorithm 1 as T-GWF algorithm for this nature.

In Line 12 of Algorithm 1, the step height is updated. In the
following ``For" loop, Before the $K_{k^*}$th time segment, the
STMs are updated with the scaled $m$th row and $m$th column of the
traffic matrix $\textbf{\emph{A}}$. The scale factor $\widetilde{\eta}$ is the ratio of
the allocated traffic for a specific participating time segment to
the total allocated traffic.

The last step of the outer ``For" loop is to allocate data when
$m=k^*$, \emph{i.e.}, for the time segment $K_{k^*}$. The data traffic  for
allocation is a sub-matrix of $\textbf{\emph{A}}$, consisting of its first
row to the $k^*$th row, and from the first column to the $k^*$th
column, as listed in Line 9. In the inner ``For" loop, the STMs are
updated in Line 20. Different with Lines 17 and 18, where each
update is carried for a row and a column respectively, each update
is adding a scaled sub-matrix.

The proposed T-GWF have two great advantages.
The first one is to distribute the data traffic  among
participating time segments to guarantee the relay being in the time windows of the involved source and destination satellites and to achieve load balance. GWF can efficiently solve this problem. The second advantage is the structure
of the water-filling solution facilitating the STMs update.

Water-filling structured solution by T-GWF also reveals that the obtained
water level for each time segment represents the corresponding
data rate, or equivalently, data traffic density. For time segment
$K_{k^*}$ to $K_S$, the water level is non-decreasing. This
reflects that satellites relaying data traffic is not uniform across the time segments, but tends to be more intense towards the central point of the time windows. This is reasonable since the closer to the central point, the more data traffic  uploaded from
the air balloons is available.

\section{Configuration Matrices Generation}
%The sub-matrices of $\textbf{\emph{A}}'$ can be determined by
%\begin{equation}
%\textbf{\emph{A}}'_v=n_0\times\textbf{\emph{A}}_v, 1\leq v\leq S.
%\end{equation}
After the STMs determination, the data traffic represented by the STM $\emph{\textbf{A}}_v$, with the dimension of $S\times S$, is relayed in the time segment of $\alpha\tau_v$.
Define a set of \emph{configuration matrices} with the dimension of $S\times S$ so that each row and each column for a configuration matrix has at most one, respectively. The rest of the elements for the configuration matrix are all zeroes.
The time for transmitting one bit of data, \emph{i.e.}, the inverse of the data transmission rate, is $\phi$ seconds, where $\phi=1/C_0$.
Assume that when $n_0=1$, each configuration matrix has a coefficient $\varphi_v$, which denotes the number of $\phi$ to relay the data traffic  in the configuration matrix.
{\color{blue}Then, for a general form when $n_0\geq1$, each configuration matrix generated from the STM $\emph{\textbf{A}}_v$ has a coefficient $n_0\varphi_v$.}
For simplicity, we assume that $\varphi_v$ is the same for each configuration matrix.
%In the same configuration matrix, the aligning manner between the transmitting-receiving antenna pairs remains as defined by the configuration matrix.

To determine the required number of active lasers per satellite, \emph{i.e.}, $m_v$, we assume that $\Phi_v$ configuration matrices are firstly generated for $\emph{\textbf{A}}_v$.
{\color{blue}If the $\Phi_v$ configuration matrices are implemented by satellites with one laser per satellite, the delay is by $\Phi_vn_0\varphi_v\phi+\Phi_v\delta $.
Then, when $m_v$ lasers are configured for each satellite, the satellites can simultaneously relay data traffic according to $m_v$ configuration matrices, which can be formulated as}
\begin{equation}
\frac{1}{m_v}\left(\Phi_vn_0\varphi_v\phi+\Phi_v\delta \right)\leq\alpha\tau_v.
\label{Shu1}
\end{equation}
In practice, $m_v$ should be small to reduce the power consumption and the hardware cost of the satellites.
We can have
\begin{equation}
\frac{\Phi_v}{m_v}\left(n_0\varphi_v\phi+\delta \right)=\alpha\tau_v.
\label{Shu2}
\end{equation}
From eq. (\ref{Shu2}), a smaller $\Phi_v$ leads to a larger $\varphi_v$. This is because that a  smaller $\Phi_v$ increases ineffective occupation time of the transmitting-receiving antenna pairs after the data traffic has been relayed in a configuration matrix. On the other hand, $\Phi_v$ should be no less than $S$ to cover all the elements in $\textbf{\emph{A}}_v$.

When $n_0=1$, we denote the average maximum summation of the elements in either a row or a column of $\textbf{\emph{A}}_v$ by $\widetilde{A}_v$.
Then, when $n_0\geq1$, the average maximum summation of the elements in either a row or a column of $\textbf{\emph{A}}_v$ is $n_0\times\widetilde{A}_v$.
We propose Theorem 1 as follows.

\underline{\textbf{Theorem 1}}  Define $\varphi_v$ as follows, with a unit of bits:
\begin{equation}
\varphi_v=\frac{n_0\widetilde{A}_v}{\Phi_v-S}, \Phi_v>S.
\label{varphi}
\end{equation}
Then, $\textbf{\emph{A}}_v$ can be covered by at most $\Phi_v$ configuration matrices.
{\color{blue}The corresponding time delay for transmitting $\varphi_v$ bits of data is $\varphi_v\phi$.}

\emph{Proof}: {\color{blue}See Appendix A.}
\hfill\rule{4pt}{8pt}

Considering Theorem 1, in eq. (\ref{Shu2}), we have
\begin{equation}
\frac{n_0\phi\widetilde{A}_v}{\Phi_v-S}+\delta\leq\alpha\tau_v, \Phi_v>S.
\label{Shuu}
\end{equation}
This guarantees the feasibility of the configuration matrices generation.

By eqs. (\ref{Shu2}) and (\ref{varphi}), we can further derive that
\begin{equation}
m_v=\frac{\Phi_v}{\alpha\tau_v}\times\left(\frac{n_0\phi\widetilde{A}_v}{\Phi_v-S}+\delta\right), \Phi_v>S.
\label{Shu3}
\end{equation}
In practice, $m_v$ can be rounded up to an integer, \emph{i.e.}, $m=\lceil m_v\rceil$.

\begin{figure}[!b]
\hrulefill
\footnotesize
\begin{equation} \tag{P2}
  \begin{split}
  &\text{minimize} \\ &\sum_{i=1}^S\frac{B_0l_i^{2}T^G_i}{n_0 G_T}\left(e^{\frac{\ln2\times n_0T^{\text{LEO}}\lambda_i}{B_0T^G_i}}-1 \right)+\sum_{i=1}^S\frac{B_1(L-l_i)^{2}T^B_i}{n_0 G_T}\left(e^{\frac{\ln2\times n_0T^{\text{LEO}}\lambda_i}{B_1T^B_i}}-1 \right)\\
  &+\sum_{i=1}^S\frac{B_2(L-l_i)^{2}T^D_i}{n_0 G_T}\left(e^{\frac{\ln2\times n_0T^{\text{LEO}}\mu_i}{B_2T^D_i}}-1 \right)+\frac{P^C\eta S}{10^{11.44}\times \sigma^2 n_0}\\
  &+\frac{\sum_{v=k^*}^S\left(\frac{\Phi_vn_0P^L_0S\widetilde{T}_{k^*}\left(\sum_{i=1}^S\sum_{j=1}^Sa_{ij}\right)\left(\frac{n_0\phi\widetilde{A}_v}{\Phi_v-S}+\delta\right)}{\left(T^Z_{k^*}\right)^2}+C_0P^L_1S\Phi_v\left(\frac{n_0\phi\widetilde{A}_v}{\Phi_v-S}+\delta\right)+P^Z\frac{\left(\Phi_v\right)^2}{T^Z_{k^*}}\left(\frac{n_0\phi\widetilde{A}_v}{\Phi_v-S}+\delta\right)S \delta_Y\right)}{10^{11.44}\times\sigma^2n_0},\\
  &\emph{s.t.} \\
  &\qquad\qquad T^G_i+\frac{\eta n_0T^{\text{LEO}}\times\sum_{i=1}^S\lambda_i }{C^{\text{HAB}}}+\frac{l_i}{V}+\widetilde{T}_i\leq T^{\text{LEO}}, 1\leq i\leq S,\\
  &\qquad \qquad T^B_i+T^D_i+2\times\frac{L-l_i}{V}+\alpha\widetilde{T}_i\leq\widetilde{T}_i, k^*\leq i\leq S, T^B_i+T^D_i+2\times\frac{L-l_i}{V}+\alpha\widetilde{T}_{k^*}\leq\widetilde{T}_i, 1\leq i< k^*,\\
 &\qquad \qquad \overline{m}=\sum_{v=k^*}^S\frac{\Phi_v}{T^Z_{k^*}}\times\left(\frac{n_0\phi\widetilde{A}_v}{\Phi_v-S}+\delta\right)\leq m_{\text{max}}, \quad \text{and}\quad \frac{n_0\phi\widetilde{A}_v}{\Phi_v-S}+\delta\leq\alpha\tau_v, k^*\leq v\leq S,\\
  &\qquad \qquad  1\leq n_0\leq n_{\text{max}}\quad \text{and}\quad   0<\alpha<1.
  \end{split}
  \label{P2}
  \end{equation}
\hrulefill
\end{figure}

\begin{figure}[!b]
\hrulefill
\footnotesize
\begin{equation} \tag{P3}
  \begin{split}
  &\text{minimize}\ F\left(\alpha, n_0, T^G_i, T^B_i, T^D_i, \{\Phi'_v\}\right)\\
  &\emph{s.t.} \\
  &\sum_{i=1}^S\sum_{t=1}^{t_{\text{max}}}\frac{B_0l_i^{2}T^G_i\left(\frac{\ln2\times n_0T^{\text{LEO}}\lambda_i}{B_0T^G_i}\right)^t}{n_0G_Tt!}
  +\sum_{i=1}^S\sum_{t=1}^{t_{\text{max}}}\frac{B_1(L-l_i)^{2}T^B_i\left(\frac{\ln2\times n_0T^{\text{LEO}}\lambda_i}{B_1T^B_i}\right)^t}{n_0G_Tt!}\\
  &+\sum_{i=1}^S\sum_{t=1}^{t_{\text{max}}}\frac{B_2(L-l_i)^{2}T^D_i\left(\frac{\ln2\times n_0T^{\text{LEO}}\mu_i}{B_2T^D_i}\right)^t}{n_0G_Tt!}+\frac{P^C\eta S}{10^{11.44}\times \sigma^2 n_0}\\
  &+\frac{\sum_{v=k^*}^S\left(\frac{S(\Phi'_v+S)n_0P^L_0\left(\sum_{i=1}^S\sum_{j=1}^Sa_{ij}\right)\left(\frac{n_0\phi\widetilde{A}_v}{\Phi'_v}+\delta\right)}{\alpha^2\widetilde{T}_{k^*}}+C_0P^L_1S(\Phi'_v+S)\left(\frac{n_0\phi\widetilde{A}_v}{\Phi'_v}+\delta\right)\right)}{10^{11.44}\times\sigma^2n_0}\\
  &+\frac{\sum_{v=k^*}^S\left(P^Z\frac{\left(\Phi'_v+S\right)^2}{ \alpha\widetilde{T}_{k^*}}\left(\frac{n_0\phi\widetilde{A}_v}{\Phi'_v}+\delta\right)S \delta_Y\right)}{10^{11.44}\times\sigma^2n_0}\leq F,\\
  &\qquad T^G_i+\frac{\eta n_0T^{\text{LEO}}\times\sum_{i=1}^S\lambda_i }{C^{\text{HAB}}}+\frac{l_i}{V}+\widetilde{T}_i\leq T^{\text{LEO}}, 1\leq i\leq S,\\
  &\qquad  T^B_i+T^D_i+2\times\frac{L-l_i}{V}+\alpha\widetilde{T}_i\leq\widetilde{T}_i, k^*\leq i\leq S,\quad\text{and}\quad
  T^B_i+T^D_i+2\times\frac{L-l_i}{V}+\alpha\widetilde{T}_{k^*}\leq\widetilde{T}_i, 1\leq i<k^*,\\
  %&\qquad \qquad
%  \sum_{v=i}^S\tau_v\leq\alpha\widetilde{T}_i,\\
  &\qquad
  \overline{m}=\sum_{v=k^*}^S\frac{\Phi_v}{\alpha \widetilde{T}_{k^*}}\times\left(\frac{n_0\phi\widetilde{A}_v}{\Phi_v-S}+\delta\right)\leq m_{\text{max}}, \quad\text{and}\quad
  \frac{n_0\phi\widetilde{A}_v}{\Phi'_v}+\delta\leq\alpha\tau_v, k^*\leq v\leq S,\\
  &\qquad
  1\leq n_0\leq n_{\text{max}}, \quad\text{and}\quad  0<\alpha<1.
  \end{split}
  \label{P3}
  \end{equation}
\hrulefill
\end{figure}

To maximize the energy efficiency of the active lasers, we define the average number of active lasers $\overline{m}$ as
\begin{equation}
\overline{m}=\frac{\alpha\sum_{v=k^*}^S m_v\tau_v}{\alpha\sum_{\widetilde{v}=k^*}^S\tau_{\widetilde{v}}}=\frac{\alpha\sum_{v=k^*}^S m_v\tau_v}{T^Z_{k^*}}
=\frac{\sum_{v=k^*}^S\Phi_v\times\left(\frac{n_0\phi\widetilde{A}_v}{\Phi_v-S}+\delta\right)}{T^Z_{k^*}}, \Phi_v>S.
\label{mmax}
\end{equation}
By eq. (\ref{EEZ}), $\varepsilon^Z$ can be re-formulated by
\begin{equation}
\varepsilon^Z=P^Z\frac{\left(\Phi_v\right)^2}{T^Z_{k^*}}\left(\frac{n_0\phi\widetilde{A}_v}{\Phi_v-S}+\delta\right)S \delta_Y.
\label{NEEZ}
\end{equation}
By eqs. (\ref{StaticLaser}) and (\ref{DynamicLaser}), the static energy consumption $\varepsilon^L_0$ and dynamic energy consumption $\varepsilon^L_1$ can be formulated in eqs. (\ref{SStaticLaser}) and (\ref{DDynamicLaser}), respectively, as follows:
\begin{equation}
\varepsilon^L_0=\frac{n_0\left(\sum_{i=1}^S\sum_{j=1}^Sa_{ij}\right)P^L_0S\widetilde{T}_k\sum_{v=k}^S\Phi_v\left(\frac{n_0\phi\widetilde{A}_v}{\Phi_v-S}+\delta\right)}{\left(T^Z_k\right)^2}.
\label{SStaticLaser}
\end{equation}
\begin{equation}
\varepsilon^L_1=C_0P^L_1S\sum_{v=k}^S\Phi_v\left(\frac{n_0\phi\widetilde{A}_v}{\Phi_v-S}+\delta\right).
\label{DDynamicLaser}
\end{equation}

%The total number of the antenna pair adjusting at a satellite is
%\begin{equation}
%\widehat{m}=\sum_{v=1}^Sm_v.
%\label{mmmax}
%\end{equation}
\begin{figure*}[!tb]
{\includegraphics[height=2.5in]{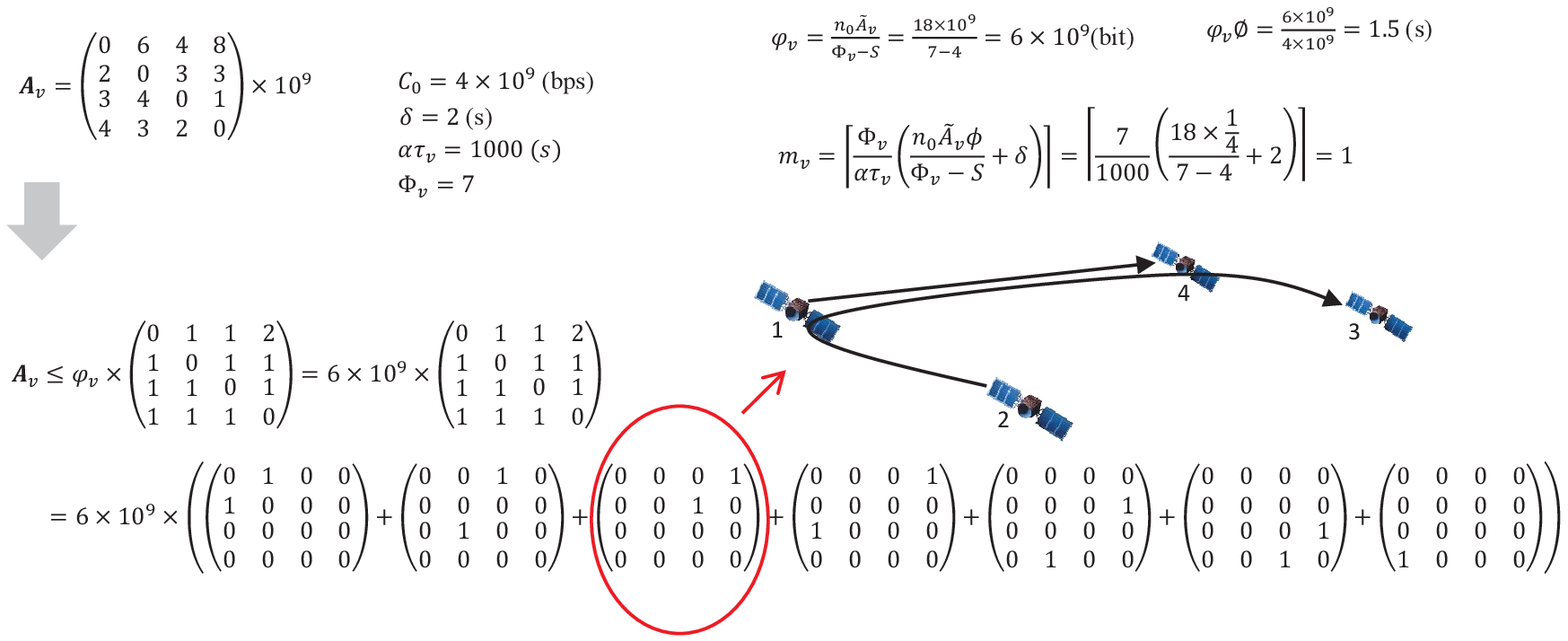}}
\renewcommand{\captionlabeldelim}{.}
\centering
\caption{An example showing the configuration matrices generation.}
\label{Routing}
\end{figure*}
In Fig. \ref{Routing},  using a LEO satellite network with 4 satellites, we demonstrate an example of the configuration matrices generation for a specific time segment \emph{v} with the sub-traffic-matrix $\textbf{\emph{A}}_v$.
%\begin{equation}\nonumber       %开始数学环境
%\textbf{\emph{A}}'_v=\left(                 %左括号
%  \begin{array}{cccc}   %该矩阵一共3列，每一列都居中放置
%    0 & 6 & 4&8\\
%    2 & 0 & 3&3\\
%    3 & 4 & 0&1\\
%    4 & 3 & 2&0\\
%  \end{array}
%\right)\times10^9.                 %右括号
%\end{equation}
{\color{blue}We assume that $C_0=4\times10^9$ bps, $\delta=2$ s, $\alpha\tau_v=1000$ s, and $\Phi_v=7$.}
{\color{blue}Then, $\phi=1/C_0=0.25\times10^{-9}$ s/bit, \emph{i.e.}, it takes $0.25\times10^{-9}$ seconds for relaying 1 bit of data among satellites}.
By Theorem 1, we can find that $\varphi_v=6\times10^9$ bits.
This suggests that the transmission delay is $\varphi_v\phi=1.5$ s for a schedule. Furthermore, $m_v$ can be calculated by eq. (\ref{Shu3}) and rounded up to an integer, which gives $m_v=1$.

$\textbf{\emph{A}}_v$ is decomposed into $\Phi_v$ configuration matrices, each of which has a coefficient $\varphi_v$.
The specific elements in a configuration matrix are determined by the traffic scheduling, which in turn depends on the data traffic routing among satellites and the number of transmitting-receiving antenna pairs, \emph{etc}.
We show an example of traffic scheduling for the $\Phi_v$ configuration matrices in Fig. \ref{Routing}.
Since $m_v=1$, each satellite can transmit at most one data stream at any given time.
Thus, it takes 7 schedules for completing the relay among satellites.
For the third configuration matrix, satellite 1 transmits to satellite 4, while satellite 2 transmits to satellite 3, using satellite 1 and satellite 4 as relays.
%Assume that the number of transmitting-receiving antenna pairs configured at each satellite is 2.
%Then, the average number of the transmitting-receiving antenna pairs in use is 1.5 per satellite in the LEO satellite network.
{\color{blue}In our future work, we will discuss the data traffic scheduling in each configuration matrix and the corresponding routing schemes for relay among satellites.}
\section{Collaborative Multi-Resource Allocation}

\textcolor{blue}{In this section, we present the approximation and transformation to solve the original optimization problem of (P1).}

\subsection{Optimization with Taylor Approximation for Collaborative Multi-Resource Allocation}
\begin{algorithm}[!tb]
\footnotesize
\caption{Algorithm for solving optimization problem (\ref{P3}).}
\label{alg:Routing}
\begin{algorithmic}[1] %这个1 表示每一行都显示数字
\renewcommand{\algorithmicrequire}{\textbf{Initialization and Input:}}
\REQUIRE ~~\\ %算法的输入参数：Input
$\{B_0, B_1, B_2\}$; $\{l_i\}$; $\sigma^2$; \emph{V}; $\{\lambda_i\}$; $\{\mu_i\}$; \emph{S}; $\delta_Y$; $P^L_0$; $P^Z$; $T^{\text{{LEO}}}$; Initialize $t_{\text{max}}=0$ and $E^{\text{Total}}=0$; Initialize $k^*=1$; Initialize $\widetilde{E}=0$.\\
\FOR{($k^*=1$, $k^*\leq S$, $k^*++$)}
\STATE
Implement Algorithm 1 to determine $\{\textbf{\emph{A}}_v\}$;
\WHILE{($E^{\text{Total}}$ has not converged) or ($t_{\text{max}}=0$)}
\STATE
$t_{\text{max}}+1\rightarrow t_{\text{max}}$;
\STATE
Let $\Phi'_v=\Phi_v-S$;
\STATE
Solve the geometric programming problem (\ref{P3}) with $t_{\text{max}}$ and $k^*$.
\STATE
Calculate $\{m_v\}$ by eq. (\ref{Shu3});
\STATE
Let $\Phi_v=\Phi'_v+S$, $\forall v$;
\STATE
Calculate $\overline{m}$ by eq. (\ref{mmax});
\STATE
Calculate $E^{\text{Total}}$ by eq. (\ref{TotalEnergy}).
\ENDWHILE
\IF{($E^{\text{Total}}>\widetilde{E}$)}
\STATE
$E^{\text{Total}}\rightarrow \widetilde{E}$;
\STATE
$k^*+1\rightarrow k^*$;
\ELSE
\STATE
break;
\ENDIF
\ENDFOR
\STATE
\textbf{Return}: $E^{\text{Total}}$. % 算法的返回值
\end{algorithmic}
\end{algorithm}
By the expansion of  the problem (\ref{P1}) with \cref{TAi,ComputingDelay,TZTZ,PTG,PTB,PTD,ComputingEnergy,NEEZ,SStaticLaser,DDynamicLaser}, as well as the constraints of $1\leq n_0\leq n_{\text{max}}$ and $\overline{m}\leq m_{\text{max}}$, optimization problem (\ref{P1}) can be transformed into problem (\ref{P2}).

Let $\Phi'_v=\Phi_v-S$.
For a variable \emph{x},
by Taylor series, we have $e^x=\sum_{t=1}^{+\infty}x^t/t!$, which leads to
\begin{equation}
2^x=e^{x\ln2}=1+\sum_{t=1}^{+\infty}\left(x\ln2\right)^t/t!.
\end{equation}
Then, optimization problem (\ref{P2}) can be further transformed into problem (\ref{P3}), which is a standard  geometric programming problem solvable by CVX \cite{Chiang}.

{\color{blue}In (\ref{P3}), $t_{\text{max}}$ is a  positive integer that denotes the maximum number of terms used in the Taylor series.}
We propose Algorithm 2 to solve problem (\ref{P3}).
In Algorithm 2, $t_{\text{max}}$ is determined in an iterative manner until the values of all optimization variables converge.
%Obviously, a larger $t_{\text{max}}$ can guarantee the problem (\ref{P2}) more approached to the problem (\ref{P3}).
{\color{blue}It is difficult to determine the optimal value of $k^*$, especially when the number of time segments is large.
In Algorithm 2, the value of $k^*$ is determined heuristically, and its value stops updating when the system energy efficiency does not increase further as $k^*$ increases.}

% After the optimization, in practice, $m_v$ should be bounded to an integer by $m_v=\lceil m_v\rceil$.
\section{Numerical Results}
\subsection{Simulations}
\begin{table}[!htbp]
\footnotesize
\captionstyle{center} \onelinecaptionsfalse
\renewcommand{\captionlabeldelim}{}
\begin{center}
\caption{\protect\\\textsc{Simulation Parameters}}\vspace{+1em}
\begin{tabular}{c|c|c|c|c|c|c|c}
\toprule
\hline
Parameter & Value&Parameter & Value&Parameter & Value&Parameter & Value\\  \hline
$t^{\text{max}}$&10&\emph{S}&5&
\emph{L}&550 km&$G_T$&$10^{\frac{15}{10}}$ (15 dB)\\
V&$3\times 10^8$ m/s&$\eta$&$10^{10}$ cycles/bit&
$C^{\text{HAB}}$&$10^{12}$ cycles/second&$C_0$&$10^9$ bps\\
$B_0$&$10^8$ Hz&$B_1$&$10^8$ Hz&
$B_2$&$10^8$ Hz&$P^A$&$10^{-10}$ W/bit\\
$P^C$&$10^{-6}$ W/cps&$P^L_0$&$10^{-15}$ W/bps&
$P^L_1$&$10^{-15}$ W/bps&$P^Z$&$10^{-3}$ W/laser\\
$\Omega$ &$2.17\times10^4$ km&$\delta_Y$&1 second&
$n_{\text{max}}$&20&$m_{\text{max}}$&50\\
\hline
\bottomrule
 \end{tabular}
  \end{center}
\end{table}

%\begin{figure}[htbp]
%\centering
%\begin{minipage}[t]{1\textwidth}
%\centering
%\includegraphics[height=1.9 in]{NNN.eps}
%\captionsetup{font={footnotesize }}
%\renewcommand{\captionlabeldelim}{.}
%\centering
%\caption{The system performance with $n_{\text{max}}$.}
%\label{NNN}
%\end{minipage}
%\vfill
%\begin{minipage}[t]{1\textwidth}
%\centering
%\includegraphics[height=1.9 in]{SSS.eps}
%\captionsetup{font={footnotesize }}
%\renewcommand{\captionlabeldelim}{.}
%\centering
%\caption{The system performance with \emph{S}.}
%\label{SSS}
%\end{minipage}
%\end{figure}

\begin{figure*}[!tb]
{\includegraphics[height=1.9in]{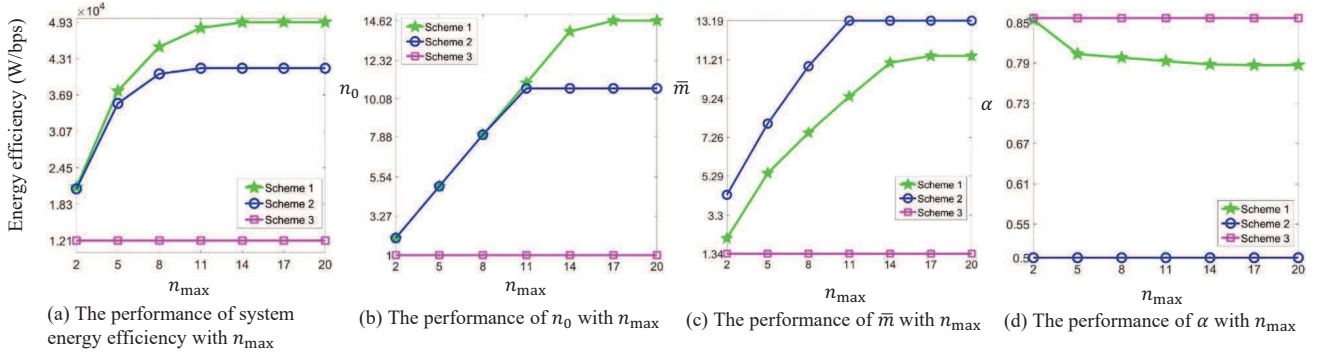}}
\renewcommand{\captionlabeldelim}{.}
\centering
\caption{The system performance with $n_{\text{max}}$.}
\label{NNN}
\end{figure*}
\begin{figure*}[!tb]
{\includegraphics[height=1.9in]{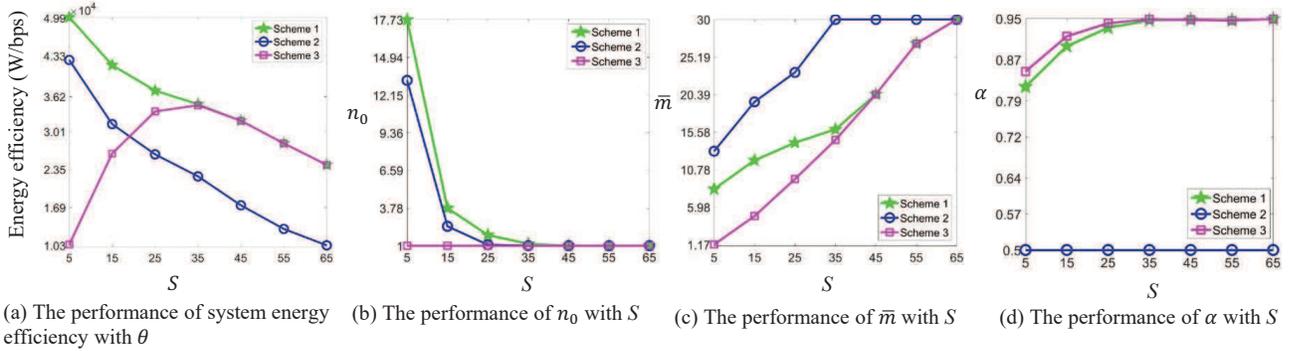}}
\renewcommand{\captionlabeldelim}{.}
\centering
\caption{The system performance with \emph{S}.}
\label{SSS}
\end{figure*}

Table I gives the general simulation parameters. The noise temperature is assumed to be 260 K.
We consider satellites deployed at the height of 550 km above the Earth, same as the setting used in the Starlink LEO network.
\textcolor{blue}{This leads to a maximum routing distance $\Omega$ being equal to half of the orbit circumference, \emph{i.e.}, $\Omega=\pi\left(L+r_E\right)=\pi\left(550+6371\right)\approx 2.17\times10^4$ km. Then, the maximum delay of data routing among satellites $\delta_R$ is $\delta_R=\Omega/V=7.25\times10^{-2}$ s.}

The hot air balloon with the minimum height is deployed at $l_{\text{min}}=20$ km, and the maximum height at $l_{\text{max}}=75$ km above the earth. The height of the remaining $S-2$ hot air balloons is evenly distributed between $l_{\text{min}}$ and  $l_{\text{max}}$. The hot air balloon at lower height generally has a larger minimum elevation to the satellite. We set the minimum elevation, for the hot air balloon at the maximum and the minimum height to the Earth, to $\beta^\text{min}=5^{\circ}$ and $\beta^\text{max}=45^{\circ}$, respectively. The minimum elevation of the other $S-2$ hot air balloons is evenly distributed between $\beta^\text{min}$ and $\beta^\text{max}$. The traffic matrix \textbf{\emph{A}} is randomly generated, where the diagonal elements are 0, and the other entries of  \textbf{\emph{A}} are random and uniformly distributed over [0, $\theta$] bits.
Unless otherwise specified, we assume that $\theta=10^4$ bits.

Since the system model that we consider is new, there is no existing algorithm that considers the same network model and scenario for comparison. To evaluate the performance of our proposed scheme, we name the scheme using both Algorithm 1 and Algorithm 2 as scheme 1.
Let scheme 2 be the semi-fixed method using Algorithm 1 and Algorithm 2 with $\alpha=0.5$ and $k^*=1$.
Let scheme 3 be the semi-fixed method using Algorithm 1 and Algorithm 2 with $n_0=1$ and $k^*=1$.
Then, the performance of scheme 1 is compared with those of scheme 2 and scheme 3.

In Fig. \ref{NNN}, the performance of the three schemes versus the maximum value of  $n_0$, \emph{i.e.}, $n_{\text{max}}$, is illustrated.
In Fig. \ref{NNN}(a), as $n_{\text{max}}$ increases, the energy efficiency of the schemes 1 and 2 firstly increases due to the increased throughput and relatively small energy consumption. As $n_{\text{max}}$ further increases, the energy efficiency gradually saturates because of the ever-increasing energy consumption.
Scheme 3 has a fixed energy efficiency performance because its $n_0$ remains 1, then the energy efficiency is independent on $n_{\text{max}}$.
Based on the figure, our proposed scheme 1 outperforms the other schemes in terms of energy efficiency.
The gap is much smaller between the performance of schemes 1 and 2 than between the performance of schemes 1 and 3. This suggests that more frequent transmission in a  terrestrial-satellite network cannot guarantee a larger energy efficiency, because the large power consumption only supports a limited amount of data traffic corresponding to a relatively small $n_{\text{max}}$.
In Fig. \ref{NNN}(b), the required $n_0$ increases for both  schemes 1 and 2. {\color{blue}This is because when the amount of data is relatively small, an increasing $n_0$ leads to a larger amount of data traffic, which enables higher energy efficiency}.
From Fig. \ref{NNN}(c), we can verify that the required $\overline{m}$ increases for schemes 1 and 2 to meet the relay requirement among satellites due to the increasing amount of data traffic.
As $n_{\text{max}}$ further increases, $\overline{m}$ remains constant for schemes 1 and 2, because the amount of data traffic  remains.
In Fig. \ref{NNN}(d),  $\alpha$ for  scheme 1 gradually decreases, because the increasing amount of data traffic  requires a longer time interval to transmit between satellites and hot air balloons.

In Fig. \ref{SSS}, the performance of the three schemes versus \textcolor{blue}{the number of satellites} \emph{S} is illustrated.
In Fig. \ref{SSS}(a), as \emph{S} increases, the energy efficiency performance of  schemes 1 and 2 monotonously decreases. This is because a larger \emph{S} results in a larger power consumption dominating the performance of energy efficiency in wireless transmission and the lasers for relay.
For scheme 3, as \emph{S} increases, the energy efficiency firstly increases because the increasing amount of data traffic  brought by the increased satellites.
As \emph{S} further increases, the energy efficiency of the schemes 1 and 3 gradually merges. This is because, as \emph{S} increases,  the optimal $n_0$ is 1 to decrease the energy consumption of lasers and wireless transmission.
This can be verified in Fig. \ref{SSS}(b), in which the required $n_0$ approaches 1 for all the schemes as \emph{S} increases.
In Fig. \ref{SSS}(b), the $n_0$ for scheme 1 is larger than that of scheme 2 when $n_{\text{max}}<25$.
This suggests that scheme 1 can support a larger amount of data traffic due to the optimal resource allocation.
In Fig. \ref{SSS}(c), as \emph{S} further increases, $\overline{m}$ also increases due to the increasing amount of data traffic. The optimal $\overline{m}$ for  scheme 2 is larger than those of schemes 1 and 3. This is caused by the smaller $\alpha$ for  scheme 2, which can be verified in Fig. \ref{SSS}(d).
In Fig. \ref{SSS}(d), as \emph{S} increases, the required $\alpha$ for both the scheme 1 and scheme 3 gradually increases, because the increased power consumption by lasers for relay among satellites.
As \emph{S} further increases, the required $\alpha$ of  schemes 1 and 3 gradually remains and merges. This is because that $n_0$ approaches 1 for both schemes. Then, the performance of scheme 1 becomes the same with that of scheme 3.

\begin{figure}[!t]
{\includegraphics[height=2in]{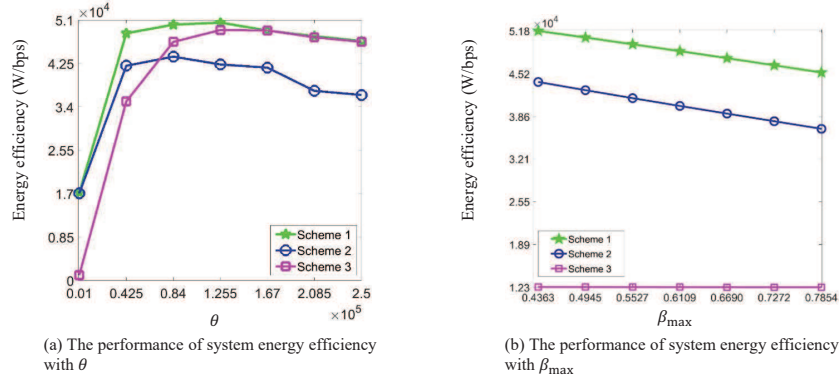}}
\renewcommand{\captionlabeldelim}{.}
\centering
\caption{The performance of the system energy efficiency with $\theta$ and $\beta_{\text{max}}$, respectively.}
\label{TY}
\end{figure}

In Fig. \ref{TY}, we illustrate the energy efficiency versus $\theta$, the maximum amount of data to relay between arbitrary two satellites, and $\beta_{\text{max}}$, the maximum minimal elevation angle, respectively.
In Fig. \ref{TY}(a), as $\theta$ increases,  the energy efficiency firstly increases for all the schemes. This is because a larger  $\theta$ can increase the data traffic  amount to improve the energy efficiency.
As $\theta$ further increases, the energy efficiency arrives the peak and then gradually falls for all three schemes, because the high power consumption used for communications.
In Fig. \ref{TY}(b), as $\beta_{\text{max}}$ increases, the energy efficiency of schemes 1 and 2 monotonously decreases. This is because a larger $\beta_{\text{max}}$  leads to a smaller time windows between hot air balloons and satellites. Accordingly, a shorter relay time among satellites causes higher power consumption of relay by lasers and reduced energy efficiency.
For the scheme 3, the energy efficiency remains constant, because a smaller $n_0$ will not cause the higher power consumption by lasers when $\beta_{\text{max}}$ is not very large.
\subsection{Discussions}
%In this work, we study the integrated energy efficiency in a terrestrial-satellite network considering the hot air balloons playing as relays with different height to the Earth.
We have studied a relatively complex system model in this work, including: 1) different hovering heights of the hot air balloons with different minimum elevation angle from hot air balloons to the satellites; 2) the caching, computing, and communication power consumption in this integrated system; 3) different lengths of time windows and the corresponding STMs determination for relay among satellites; 4) the configuration matrix generation to form multiple configuration matrices corresponding to the schedules; 5) an integrated optimization of the variables to maximize the overall system energy efficiency.

Two directions can be considered for future works, which can advance the research in terrestrial-satellite networks.

Firstly, joint traffic scheduling and routing scheme is of interest. As illustrated in Fig. \ref{Routing}, the joint traffic scheduling and routing can  be considered to determine the elements in a specific configuration matrix. The elements in configuration matrices will determine the number of the active lasers and affect the number of transmitting-receiving antenna pairs.
Additionally, the routing scheme of traffic relay among satellites also effects the performance of the total routing distance and the number of transmitting-receiving antenna pairs in use at each satellite.

\begin{figure}[!t]
	{\includegraphics[height=2.1in]{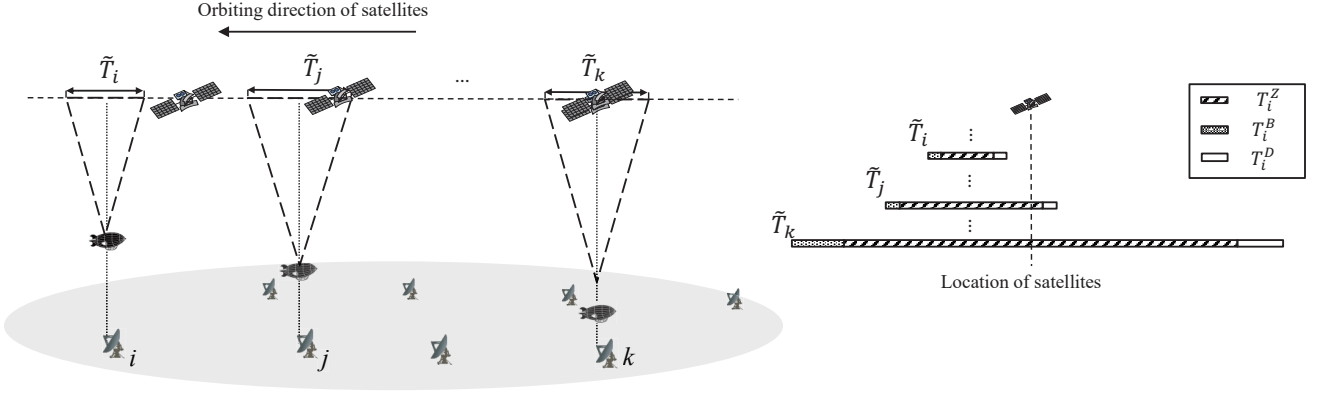}}
	\renewcommand{\captionlabeldelim}{.}
	\centering
	\caption{A more complex relationship among time windows.}
	\label{ComplexTW}
\end{figure}

{\color{blue}Secondly, in this work, we assume that the time windows are overlapping and symmetrical with respective to a central line, as shown in the lower part of Fig. \ref{fig:Topology}. The central line represents the assumption that the satellites will simultaneously arrive right above their corresponding hot air balloons. For a large satellite network, such assumption might not be practical. Instead, as shown on the left-hand side of Fig. \ref{ComplexTW}, the starting point of the time windows for the satellites may differ with each other, leading to asymmetric time windows as shown on the right-hand side of Fig. \ref{ComplexTW}.
In the figure, when the satellite \emph{k} is right above its hot air balloon, the time window of satellite \emph{j} just starts, and the time window of satellite \emph{i} is yet to start.
In this scenario, the STM determination needs further investigation.}

{\color{blue}Lastly, applications, such as content provision \cite{Zhang20_TMC}, can be studied based on our proposed framework for TSN.}

\section{Conclusion}

In this work, we studied the energy efficiency performance of TSN while jointly considering the caching, computing, and communication resources.
We considered a practical scenario, in which different hot air balloons can be deployed at different heights with different minimum elevation angles.
To effectively utilize the time windows between satellites and hot air balloons,  we investigated STM determination according to the length of the time windows, through which we achieved balanced load for relaying and guaranteed that relaying happens within the time windows of the corresponding source and target satellites.
We also proposed a configuration matrix generation algorithm to obtain the optimal number of lasers per satellite under the constraint of the available relay time.
Then, we solved the collaborative multi-resource allocation problem to optimize the transmission power, the serving period of satellites, and the required number of lasers per satellites at the relays for the maximum system energy efficiency. Simulation results verified the effectiveness of our proposed scheme, and potential future directions were discussed.

\begin{appendices}
% \clearpage % 另起一页
\section{Proof of Theorem 1}
\emph{Proof}: since the maximum line summation of $\textbf{\emph{A}}_v$ is $n_0\widetilde{A}_v$, \emph{i.e.}, $n_0\times\left(\sum_{i=1}^Sa_{ij}^v\right)\leq n_0\widetilde{A}_v$ and $n_0\times\left(\sum_{j=1}^Sa_{ij}^v\right)\leq n_0\widetilde{A}_v$.
Define $\textbf{\emph{Q}}=\{q_{ij}\}$ and $\textbf{\emph{R}}=\{r_{ij}\}$,   Let
\begin{equation}
\textbf{\emph{A}}_v=\frac{n_0\widetilde{A}_v}{\Phi_v-S}\times \emph{\textbf{Q}}+\emph{\textbf{R}},
\label{QRM}
\end{equation}
where
\begin{equation}
\sum_{i=1}^Sq_{ij}=\sum_{i=1}^S\left\lfloor\frac{n_0a_{ij}^v}{n_0\widetilde{A}_v/\left(\Phi_v-S\right)}\right\rfloor
\leq\left\lfloor\frac{n_0\times\left(\sum_{i=1}^Sa_{ij}^v\right)}{n_0\widetilde{A}_v/\left(\Phi_v-S\right)}\right\rfloor
\leq\left\lfloor\frac{n_0\widetilde{A}_v}{n_0\widetilde{A}_v/\left(\Phi_v-S\right)}\right\rfloor\leq\Phi_v-S,
\end{equation}
and $\sum_{j=1}^Sq_{ij}\leq\Phi_v-S$.

Hence, the maximum line summation of $\textbf{\emph{A}}_v$ is $\Phi_v-S$.
According to graph theory \cite{GRAPH}, the corresponding bipartite graph of \textbf{\emph{Q}} has the maximum endpoint degree $\Phi_v-S$.
Then, \textbf{\emph{Q}} can be decomposed into $\Phi_v-S$ configuration matrices with the coefficient of 1 for each.
Moreover, by eq. (\ref{QRM}), we have $r_{ij}<n_0\widetilde{A}_v/(\Phi_v-S)$.
%\begin{equation}
%r_{ij}<\frac{\widetilde{A}'_v}{\Phi_v-S}.
%\end{equation}
This suggests that \textbf{\emph{R}} can be covered by at most \emph{S} configuration matrices with the coefficient of $n_0\widetilde{A}_v/\left(\Phi_v-S\right)$ for each.
Then, $\textbf{\emph{A}}_v$ can be covered by at most $\Phi_v$ configuration matrices, each of which the coefficient of $\varphi_v=n_0\widetilde{A}_v/\left(\Phi_v-S\right)$.
\hfill\rule{4pt}{8pt}

%\section{Proof of Theorem 2}
%\emph{Proof}:
%the standard form of geometric programming \cite{Chiang} is
%\begin{equation}\nonumber
%  \begin{split}
%  &\text{minimize}\ \{f_0(\textbf{\emph{x}})\},\\
%  &\emph{s.t.} \qquad f_i(\textbf{\emph{x}})\leq1, i=1, 2, \ldots, I,\\
%  & \qquad \quad h_j(\textbf{\emph{x}})=1, j=1, 2, \ldots, J,\\
%  \end{split}
%\end{equation}
%where $\{f_i\}$ are posynomials, each of which is defined as the summation of multiple monomials, and $\{h_j\}$ ($h_j>0$) are monomials.
%It can be easily proved that the optimization problem of (\ref{P3}) is a geometric programming problem in the standard form.
%
%\hfill\rule{4pt}{8pt}
\end{appendices}

\footnotesize
\bibliographystyle{IEEEtran}
\bibliography{Ref}

\end{document}